\documentclass[12pt,letterpaper]{article}
\usepackage[utf8]{inputenc}

\usepackage{graphicx,array}
\usepackage{url}
\usepackage{color}
\usepackage{latexsym}
\usepackage{amsthm}
\usepackage{amsmath}
\usepackage{amssymb}
\usepackage{amsfonts}
\usepackage[numbers,sort&compress]{natbib}
\usepackage{bm}
\usepackage{slashed}
\usepackage{mathrsfs}
\usepackage{enumerate}
\usepackage{tikz}
\usepackage{siunitx}
\usepackage{mdframed}
\usepackage{setspace}  
\usepackage{esvect}

\usepackage{tcolorbox}%

\usepackage{hyperref} 
\hypersetup{
    colorlinks=true,       
    linkcolor=red,          
    citecolor=blue,        
    filecolor=magenta,      
    urlcolor=blue           
}
\usepackage[all]{hypcap} 
\usepackage{multirow}
\usepackage{multicol}

\usepackage{natbib}
\newcommand{\nc}{\newcommand}  
\newcommand{\mc}{\mathcal}
\newcommand{\bv}[1]{{\bf #1}}
\newcommand{\uu}{\;\!}

\nc{\beq}{\begin{equation}}
\nc{\eeq}{\end{equation}}
\nc{\beqa}{\begin{eqnarray}}  
\nc{\eeqa}{\end{eqnarray}}  
\nc{\bit}{\begin{itemize}}  
\nc{\eit}{\end{itemize}}  

\def\GeV{\mathrm{GeV}}     

\newcommand{\eg}{{\it e.g.}}
\newcommand{\ie}{{\it i.e.}}

\newcommand{\mmed}{m_{\rm med}}
\newcommand{\GpoG}{\beta}
\newcommand{\fGW}{f_{\rm GW}}
\newcommand{\fGWs}{f_{{\rm GW},s}}
\newcommand{\MDM}{\text{{\bf \tiny MDM}}}
\newcommand{\mdm}{M_\MDM}
\newcommand{\rdm}{R_\MDM}
\newcommand{\rhodm}{\rho_\MDM}
\newcommand{\mtot}{m}
\newcommand{\DNeff}{\Delta N_{\rm eff}}
\newcommand{\EM}{{\rm DF}}

\title{ 
 {\bf Gravitational Waves From More Attractive Dark Binaries
 }\\
\author{\large Yang Bai$^{\,a}$, Sida Lu$^{\,b}$, and Nicholas Orlofsky$^{\,c}$}
\date{\small \it 
$^a$Department of Physics, University of Wisconsin-Madison, Madison, WI 53706, USA\\
$^b$Institute for Advanced Study, The Hong Kong University of Science and Technology, \\Clear Water Bay, Kowloon, Hong Kong S.A.R., P. R. China \\
$^c$Institute of Theoretical Physics, Faculty of Physics, University of Warsaw, ul. Pasteura 5, PL-02-093 Warsaw, Poland \\
}
}

\begin{document}

\maketitle

\setlength{\parskip}{0.2ex}

\begin{abstract}
The detection of gravitational waves (GWs) has led to a deeper understanding of binaries of ordinary astrophysical objects, including neutron stars and black holes. In this work, we point out that binary systems may also exist in a dark sector with astrophysical-mass macroscopic dark matter. These ``dark binaries," when coupled to an additional attractive long-range dark force, may generate a stochastic gravitational wave background (SGWB) with a characteristic spectrum different from ordinary binaries. We find that the SGWB from planet-mass dark binaries is detectable by space- and ground-based GW observatories. The contribution to the SGWB today is smaller from binaries that merge before recombination than after, avoiding constraints on extra radiation degrees of freedom while potentially leaving a detectable GW signal at high frequencies up to tens of GHz.
\end{abstract}

\thispagestyle{empty}  
\newpage   
\setcounter{page}{1}  

\begingroup
\hypersetup{linkcolor=black,linktocpage}
\tableofcontents
\endgroup

\newpage

\section{Introduction}

The recent evidence of a stochastic gravitational wave background (SGWB) using pulsar timing arrays (PTAs) from NANOGrav~\cite{NANOGrav:2023gor,NANOGrav:2023hde,NANOGrav:2023hvm,NANOGrav:2023hfp}, EPTA and InPTA~\cite{EPTA:2023sfo,EPTA:2023fyk,EPTA:2023gyr,EPTA:2023xxk}, PPTA~\cite{Reardon:2023gzh,Reardon:2023zen,Zic:2023gta}, and CPTA~\cite{Xu:2023wog} following the first direct detection of gravitational waves (GWs)~\cite{LIGOScientific:2016aoc} at LIGO \cite{LIGOScientific:2014pky} offers an entirely new way to measure the properties and contents of the Universe.  
GWs have already provided a window into the mergers of both stellar-mass black holes (BHs) and neutron stars~\cite{LIGOScientific:2018mvr,LIGOScientific:2020ibl,KAGRA:2021vkt}, and the SGWB also gives insight into supermassive BHs~\cite{NANOGrav:2023hfp,EPTA:2023xxk}.
With other observatories like LISA~\cite{LISA:2017pwj} and SKA~\cite{Carilli:2004nx} in development, as well as several proposed observatories under active consideration~\cite{Corbin:2005ny,Reitze:2019iox,Punturo:2010zz,Maggiore:2019uih,Seto:2001qf,Hu:2017mde},  the mystery of GWs over a larger frequency band will soon be unveiled with good precision.

Besides employing GWs for understanding astrophysical objects, one can also harness GWs, especially the SGWB, to investigate early-universe physics within or beyond the Standard Model (BSM). Due to the graviton's ultra-weak interactions with other particles in the early universe, GWs serve as an excellent ``cosmic envoy," potentially containing information about the Universe's history as early as the inflation era. For example, first-order phase transitions at the QCD and/or electroweak scales might produce SGWBs of sufficient magnitude to be observable with the LISA telescope (see Ref.~\cite{Athron:2023xlk} for a recent review and Refs.~\cite{NANOGrav:2023hvm,EPTA:2023xxk,Ellis:2023oxs} for more general discussion in light of the first SGWB signal). Unlike SGWB sources generated in nonequilibrium environments, this work focuses on a novel source of SGWB from BSM: ``dark binaries" or binaries comprised of two macroscopic dark matter (MDM) objects.

The SGWB signal from the mergers of MDM objects such as primordial BHs has been a topic of recent interest \cite{Mandic:2016lcn,Wang:2016ana,Raidal:2017mfl,Sasaki:2018dmp,Pujolas:2021yaw,Garcia-Bellido:2021jlq,Atal:2022zux,Braglia:2022icu,Banerjee:2023brn}. A majority of MDM binaries are expected to form in the early universe, before matter-radiation equality \cite{Nakamura:1997sm,Ioka:1998nz,Sasaki:2016jop,Ali-Haimoud:2017rtz,Raidal:2017mfl,Raidal:2018bbj,Vaskonen:2019jpv,Hutsi:2020sol}. These binaries then inspiral and merge throughout cosmic time, and the GWs they emit contribute to the SGWB. The SGWB signal is fundamentally limited by the fact that gravity controls all the dynamics of the binary system, determining its orbital parameters and emission of energy. Thus, for example, very light MDM will not produce a detectable signal because the  amplitude of the emitted GW spectrum is directly related to the mass.

However, MDM need not only couple to gravity. In general, the dark sector may contain a variety of fields and forces, much like the visible sector. If a long-range dark force (DF) exists that couples solely to dark sector states, it can modify the SGWB signal in a number of ways. For concreteness, consider an attractive DF. First, the orbital dynamics are modified. With an attractive DF, the orbital frequency increases, leading to GW emission with both higher amplitude and frequency. Second, the binary binding energy is altered, allowing more energy to be released as radiation. Third, the MDM binary can emit the DF mediator as radiation in addition to GWs. While this does remove some energy that could be emitted as GWs, it also shortens the binary merger time (for fixed initial orbital parameters), which can enhance the merger rate. Fourth and finally, a pair of MDM objects can more easily find each other in the early universe, allowing them to form earlier and with smaller initial separation. This tends to enhance the merger rate and, consequently, the SGWB. However, in more extreme cases, the binaries may merge too early, resulting in the redshifting away of their GWs.

This work evaluates all of these effects. In particular, we demonstrate that the inclusion of the DF can enhance the SGWB from MDM binaries, rendering it more detectable for GW observatories across the frequency spectrum---from nHz at PTAs up through kHz at terrestrial interferometers like LIGO. DF-mediated MDM binaries may even present a novel target for high-frequency GW detectors. 
The binary-sourced SGWB can be sufficiently small during recombination to evade constraints on additional dark radiation degrees of freedom, but later grow larger as more binaries merge, potentially becoming detectable. Related work exploring the inclusion of additional forces for binaries in the visible rather than dark sector can be found in Refs.~\cite{Krause:1994ar,Alexander:2018qzg,Dror:2019uea}.

Note that MDMs may make up either some or all of dark matter (DM). For masses above $10^{23}~\text{g}$ up to order solar mass, they are generally constrained by microlensing to be less than about $10^{-1}$ to $10^{-2}$ of DM, depending on their mass and radius~\cite{Niikura:2017zjd,Smyth:2019whb,Griest:2013aaa,Macho:2000nvd,EROS-2:2006ryy,Wyrzykowski_2011,Niikura:2019kqi,Zumalacarregui:2017qqd,Oguri:2017ock,Croon:2020wpr,Bai:2020jfm}. Future searches may extend to smaller (currently unconstrained) masses~\cite{Katz:2018zrn,Bai:2018bej,Jung:2019fcs}. For masses above solar mass, their abundance is even more strongly constrained by accretion~\cite{Ricotti:2007au,Ali-Haimoud:2016mbv,Poulin:2017bwe,Serpico:2020ehh,Bai:2020jfm}, star survival in dwarf galaxies~\cite{Brandt:2016aco,Koushiappas:2017chw}, and Lyman-$\alpha$~\cite{Afshordi:2003zb,Murgia:2019duy}. Constraints on a long-range DF also exist, but they are generally fairly weak if the range is less than of order tens of kiloparsecs \cite{Savastano:2019zpr,Bogorad:2023wzn}. This work assumes that the DF mediator is approximately massless in comparison to all other scales in the problem while remaining compliant with existing bounds, with the case of a more massive mediator briefly discussed in the last section but ultimately left for future work.

This work is organized as follows. Sec.~\ref{sec:model} discusses models for MDM with a long-range DF. Readers who are less interested in models and more interested in phenomenology may skip to Sec.~\ref{sec:EM_emissions}, which introduces a model-independent treatment of the radiation of both GWs and DF mediators from binaries in the presence of the new DF. Sec.~\ref{sec:rate-SGWB} discusses the calculation of the binary formation and subsequent merger rate, and then combines this with the results of the Sec.~\ref{sec:EM_emissions} to calculate the SGWB spectrum. Sec.~\ref{sec:results} analyzes the SGWB spectrum both analytically and numerically and compares against existing and future GW observatory sensitives. Finally, discussion and conclusions are presented in Sec.~\ref{sec:conclusion}.

\section{Macroscopic dark matter with a dark long-range force}
\label{sec:model}

Independent of the ultralight DF mediator, MDM is assumed to form at an early time in the cosmological history. Among many macroscopic dark matter models, we choose the ``dark quark nugget" (DQN)~\cite{Bai:2018dxf} as a representative (see also \cite{Krnjaic:2014xza,Bai:2018vik,Liang:2016tqc}). In this class of models, the DQNs are made of dark quarks that are in the deconfined phase of the underlying dark quantum chromodynamics (dQCD), with the Lagrangian as 
\beqa
\mathscr{L}_{\rm dQCD} = \sum^{N_f}_{i = 1} \left[ \bar{\psi}_i i \gamma^\mu D_\mu \psi_i - m_{\psi_i}\,\bar{\psi}_i \psi_i\right] \, - \, \frac{1}{4} G^a_{\mu\nu} G^{\mu\nu \,a} ~,
\eeqa
Here, $N_f$ is the number of flavors of dark quarks $\psi_i$ with $i=1,2,\cdots,N_f$; $D_\mu \psi_i \equiv \partial_\mu \psi_i - i\,g_d\,G_\mu^a T^a \psi_i$ with $T^a$ ($a=1,2,\cdots,N_d^2-1$) as the $SU(N_d)$ generators;  $g_d$ is the dQCD gauge coupling; and $G^a_{\mu\nu} = \partial_\mu G^a_\nu - \partial_\nu G^a_\mu + g_d \, f^{abc}\,G^b_\mu G^c_\nu$ is the dark gluon field tensor. Similar to QCD in the Standard Model (SM), the dQCD gauge coupling becomes nonperturbative at around the dQCD confinement scale $\Lambda_d$.

Similar to white dwarfs and at approximately zero temperature, the DQNs have the degenerate Fermi pressure of the dark quarks balanced by the vacuum pressure $B = \Delta P_{\rm vacuum} = P_{\rm confined} - P_{\rm unconfined} \approx  \Lambda_d^4$. In the light dark quark mass limit with $m_{\psi_i} \ll \Lambda_d$, the energy density of DQNs is related to the ``bag parameter" $B$ by $\rho = 4\,B$, which provides a relation between the DQN radius and mass 
\beqa
R \approx \left(800\,\mbox{m}\right)\, \left( \dfrac{M}{M_\odot} \right)^{1/3}\, \left( \dfrac{1\,\mbox{GeV}}{\Lambda_d} \right)^{4/3} ~. 
\eeqa
In the above formula we have ignored the gravitational pressure and energy, which is satisfied when $G M \lesssim R$ or when $M \lesssim 0.4\,M_\odot \,(1\,\mbox{GeV}/\Lambda_d)^2$. The mass per baryon of a DQN is $3^{1/4}\sqrt{2\pi}\,N_d^{3/4}\,N_f^{-1/4}\,\Lambda_d$, which could be lighter than a free dark baryon with a mass proportional to $N_d\,\Lambda_d$ in the large $N_d$ limit.

On top of the dQCD interaction for the dark quarks, we also introduce an additional long-range DF between two dark (anti-)quarks. For a scalar mediator $\phi$, dark quarks have the following Yukawa-type interaction:
\beqa\label{eq:yukawa}
\mathscr{L}_{\rm Yukawa} = \frac{1}{2} \partial_\mu \phi \partial^\mu \phi - \sum_i (m_{\psi_i}\,+\, y_i\,\phi)\,\bar{\psi}_i\,\psi_i \,-\, V_0(\phi) ~,
\eeqa
with the potential $V_0(\phi)$ containing the mediator mass term $\frac{1}{2}\mmed^2 \phi^2$ as well as other self-interacting terms like $\frac{1}{4}\lambda \phi^4$, which is ignored in this study by assuming a small $\lambda$.  Further, we consider the ultralight mediator mass regime with the range of the force $1/\mmed$ much longer than the DQN radius $R$, or $1/\mmed \gg R$, such that two isolated DQNs could still interact with each other by exchanging $\phi$.

We note that the dark quark finite density could modify the vacuum expectation value of the $\phi$ field inside the DQNs as well as provide a ``plasma mass" for the scalar field to suppress the self-potential energy mediated by the scalar force. Including the fermion effective mass, the fermion finite density generates the following effective scalar potential:
\beqa
V_{1} = \mathfrak{g}\, \frac{1}{2\pi^2}\, \int^{k_F}_0 dk\,k^2\,\sqrt{k^2 \,+\, (m_\psi + y\, \phi)^2} ~,
\eeqa
where $\mathfrak{g} = 2 N_d N_f$ accounts for dark quark spin, color, and flavor; $m_{\psi_i} = m_{\psi}$ and $y_i = y$ for simplicity. The Fermi momentum is $k_F^2 = \mu^2 -  (m_\psi + y\, \phi)^2$. Working in the limit of $|m_\psi + y \phi| \ll \mu$, the above integration is approximately 
\beqa
\label{eq:potential-V1}
V_{1} \approx \frac{\mathfrak{g}}{8\pi^2}\,\mu^2\,\left[\mu^2 - (m_\psi + y\, \phi)^2 \right] ~.
\eeqa
The leading term multiplied by a factor of 1/3 provides the leading degenerate Fermi pressure for relativistic fermions. Its balance with the vacuum pressure determines the equilibrium value for the chemical potential as $\mu_{\rm eq} = (24\pi^2\,B/\mathfrak{g})^{1/4}$, which can be used to determine the number density as $n_\psi = \frac{\mathfrak{g}}{6\pi^2}\,\mu^3_{\rm eq}$ and the total number of dark quarks as $N_\psi = n_\psi (4\pi R^3)/3$. The term linear in $\phi$, proportional to $y\,m_\psi$, provides a charge density to emit the $\phi$ field. In the limit of $m_\psi = 0$, the dark quarks are in the fully-occupied chiral fermion states below the Fermi momentum. The emission of a $\phi$ particle that couples to both left-handed and right-handed fermions is Pauli-blocked.

Note that the scalar effective mass $m_{\rm in}^2 = \mathfrak{g}\,y^2/(4\pi^2)\,\mu_{\rm eq}^2$ inside the DQN could be much larger than the free scalar mass $\mmed$. 
An effective mediator mass inside the DQN larger than $1/R$ will lead to an additional reduction of $1/(m_{\rm in} R)^2$ for the effective $\phi$-charge of the DQN as a result of the screening from the dark quark plasma. This is detailed in Appendix~\ref{app:scalar-model-effective-charge}, where we solve the scalar classical equation of motion with both the source term [the linear term in $\phi$ in (\ref{eq:potential-V1})] and the location-dependent effective mass term.
Therefore, we consider the parameter region with $|m_{\rm in}| R < 1$, or 
\beqa
\label{eq:constrain-y}
y < 2^{19/12}3^{-7/12}\pi^{5/6}\mathfrak{g}^{-1/4}\,\Lambda_d^{1/3}\,M^{-1/3} = (2\times 10^{-19})\,\left(\frac{\Lambda_d}{1\,\mbox{GeV}} \right)^{1/3}\,\left(\frac{M_{\odot}}{M}\right)^{1/3}\,\left(\frac{\mathfrak{g}}{18}\right)^{-1/4} ~,
\eeqa
where $\mathfrak{g} = 2\times 3 \times 3$ was chosen similar to ordinary QCD.

The effective ``charge" of the DQN coupling to the scalar field is $q_{\rm eff} y =$ $\frac{\mathfrak{g}}{4\pi^2}\,m_\psi\,\mu^2_{\rm eq}\, y \, (\frac{4\pi}{3}\,R^3) =$ $\frac{3}{2}\,y\,\frac{m_\psi}{\mu_{\rm eq}}\,N_\psi$, from which the ratio of the “dark force” over the gravitational force between two DQNs is independent of the DQN total masses and is
\beqa\label{eq:alpha_finite_density}
\alpha = \frac{3\,\mathfrak{g}\,y^2\,m_\psi^2}{128\pi^3\,G\,\Lambda_d^4}\lesssim (0.02)\left(\dfrac{m_\psi/\Lambda_d}{0.5}\right)^2\left(\dfrac{1\uu\GeV}{\Lambda_d}\right)^{4/3}\left(\dfrac{M_\odot}{M}\right)^{2/3} \,\left(\frac{\mathfrak{g}}{18}\right)^{1/2} ~,
\eeqa
for $y$ satisfying \eqref{eq:constrain-y}. 
If the condition $|m_{\rm in}| R < 1$ is not satisfied, one could take into account of the additional charge suppression factor of $1/(|m_{\rm in}| R)^2$ and have the ratio  as 
\beqa
\alpha = \frac{\left(\frac{\pi}{3}\right)^{1/3}\,m_\psi^2}{3\times 2^{2/3}\,y^2\,G\,\Lambda_d^{8/3}\,M^{4/3}} \lesssim  (0.02)\left(\dfrac{m_\psi/\Lambda_d}{0.5}\right)^2\left(\dfrac{1\uu\GeV}{\Lambda_d}\right)^{4/3}\left(\dfrac{M_\odot}{M}\right)^{2/3} \,\left(\frac{\mathfrak{g}}{18}\right)^{-1/2} ~,
\eeqa
after using the opposite limit for $y$ in \eqref{eq:constrain-y}. 
In general, the long-range dark force becomes stronger for a smaller dark confinement scale $\Lambda_d$ and a lighter DQN mass. 

Since the scalar-mediated forces between two fermions, two anti-fermions, or one fermion and one anti-fermion are all attractive, the forces between any DQNs are attractive, regardless whether they are DQNs or anti-DQNs. Because scalar-exchanged long-range forces are coherent, one needs to require the scalar-exchanged potential energy be less than the total mass so as not to change the general features of DQNs compared to the same model without an additional long-range force. The constraint is $y < 4\times (2/3)^{5/6}5^{1/2}\pi^{4/3}\mathfrak{g}^{-1/2}\,(\Lambda_d/m_\psi)(\Lambda_d/M)^{1/3}$, which is less stringent than the constraint in \eqref{eq:constrain-y}.

For the simplified case with the same dark quark masses and Yukawa couplings $m_{\psi_i} = m_\psi$ and $y_i=y$ for all $i$, the ratio of the effective charge $q_{\rm eff}$ over the DQN mass $M$ is fixed and independent of $M$ in the limit of $|m_{\rm in}| R < 1$. For a more general case with different values of $m_{\psi_i}$ and $y_i$, there is a global $U(1)^{N_f}$ flavor symmetry. For a dark quark nugget with total $N_\psi$ dark quark number, the number of dark quarks of each flavor $i$ can be labelled by $(N_{\psi_1}, N_{\psi_2},\cdots,N_{\psi_{N_f}})$ with $\sum_{i=1}^{N_f} N_{\psi_i} = N_\psi$. The effective charge to emit a $\phi$ particle is $q_{\rm eff}y_{(N_{\psi_1}, N_{\psi_2},\cdots,N_{\psi_{N_f}})} = \frac{3}{2}\,\sum_{i=1}^{N_f} y_i\,m_{\psi_i}\,N_{\psi_i}/\mu_{\rm eq}$. The DQN mass is $M \approx 2^{3/4} 3^{1/4}\pi^{1/2}\mathfrak{g}^{-1/4}\Lambda_d\,N_\psi$ by ignoring the bare dark quark mass contributions. The ratio of the effective charge over mass is 
\beqa
\label{eq:scalar-charge-mass-ratio}
\frac{q_{\rm eff}\,y}{M}(N_{\psi_1}, N_{\psi_2},\cdots,N_{\psi_{N_f}}) = \frac{\sqrt{\frac{3}{2}}\sqrt{\mathfrak{g}}}{4\pi\,\Lambda_d^2}\, \frac{\sum_{i=1}^{N_f} y_i\,m_{\psi_i}\,N_{\psi_i}}{\sum_{i=1}^{N_f} N_{\psi_i}} ~.
\eeqa
So, different flavor-charged dark quark nuggets could have different charge-to-mass ratios. 

One formation mechanism of dark quark nuggets has been studied in Ref.~\cite{Bai:2018dxf} based on the first-order phase transition in the dark QCD sector (see also \cite{Hong:2020est,Lu:2022paj,Kawana:2022lba,Bai:2022kxq}). With an initial asymmetric dark baryon number, the average dark baryon number for each nugget is approximately the total number of dark baryons divided by the total number of bubbles within a Hubble volume at the time of phase transition. The existence of the super weak dark force does not affect this formation mechanism. Additionally, depending on the mass difference for the baryons between the true and false vacua, most of the dark quarks may end up either inside or outside of the DQNs. Thus DQNs may make up nearly all of DM, or only a small fraction with free dark baryons dominating the DM energy density.

The second class of models with fermion constituents has a vector boson force mediator, similar to the photon for the electromagnetic force. The Lagrangian contains the mediator mass and interaction terms
\beqa \label{eq:Lvector}
\mathscr{L}_{\rm vector}  \supset  \frac{1}{2}\,\mmed^2\, V_\mu V^\mu + g\, \sum_i\,\mathfrak{q}_i\,V_\mu \bar{\psi}_i \gamma^\mu \psi_i ~.
\eeqa
The mediator mass $\mmed$ is much lighter than $1/R$. For simplicity, we assume all dark quarks have the same charge $\mathfrak{q}_i = 1$ under the Abelian gauge symmetry and ignore the dark quark masses for this model. Similar to the scalar mediator case, the Fermi finite density generates a screening effect for the static Coulomb charge. The ``Debye length" is $\lambda_{\rm D} = \sqrt{\pi/(2\,\mathfrak{g}\,g^2)}\mu_{\rm eq}^{-1}$~\cite{Jancovici}. Requiring the Debye length to be longer than the size of the DQN, one has 
\beqa
\label{eq:ed-upper-limit}
g < 2^{1/12}3^{-7/12}\pi^{1/3}\mathfrak{g}^{-1/4}\,\Lambda_d^{1/3}\,M^{-1/3} = (4\times 10^{-20})\,\left(\frac{\Lambda_d}{1\,\mbox{GeV}} \right)^{1/3}\,\left(\frac{M_{\odot}}{M}\right)^{1/3} ~,
\eeqa
with $\mathfrak{g}=2\times 3 \times 3$. The ratio of the dark long-range force over the gravitational force is 
\beqa \label{eq:alpha-upper-vector}
\alpha \equiv \frac{\frac{g^2}{4\pi}\,N_{\psi\,1}N_{\psi\,2}}{G\,M_1\,M_2} = \frac{\mathfrak{g}^{1/2}\,g^2}{8\sqrt{6}\pi^2\,G\,\Lambda_d^2} < (4.6\times 10^{-3}) \, \left(\frac{1\,\mbox{GeV}}{\Lambda_d}\right)^{4/3}\,\left(\frac{M_{\odot}}{M}\right)^{2/3} ~. 
\eeqa
The self-interaction potential energy from the Abelian gauge force is less than $M/(60\pi)$ using the constraint in \eqref{eq:ed-upper-limit} and can be neglected. 

Different from the dark scalar force case, the dark gauge charge case is anticipated to produce DQNs with an average gauge charge of zero. Therefore, the formation mechanism from a first-order phase transition leads to a typical nugget net charge due to Poisson fluctuations given by the square root of the average number of dark quarks and anti-quarks in a volume determined by the bubble nucleation separation. This results in a smaller charge-over-mass ratio compared to the case of an initial charge asymmetry as was possible for the scalar charge. However, there could exist also dark electrons that are not charged under dQCD but are charged under the Abelian gauge group. The dark electrons can neutralize the Abelian gauge charge of dark baryons, similar to the situation of the ordinary electrons and protons. Depending on the dark electron mass and gauge coupling, the Bohr radius of the dark electron could be much longer than the dark quark nugget DQN size. Further, the ``recombination" temperature of dark electrons could be much lower than the relevant temperature when dark binaries emit the SGWB. With the conditions in the previous two sentences, the DQN can have a larger charge-over-mass ratio during their mergers, but the DF between two DQNs is repulsive and thus less interesting for producing an enhanced SGWB signal.

Other macroscopic DM models that may admit scalar or vector mediators in a similar way include Q-balls \cite{Rosen:1968mfz,Friedberg:1976me,Coleman:1985ki} (including gauged Q-balls \cite{Lee:1988ag,Gulamov:2015fya,Brihaye:2015veu,Heeck:2021zvk} or Q-monopole-balls \cite{Bai:2021mzu}); fermion solitons/fermi-balls \cite{Macpherson:1994wf,Hong:2020est,DelGrosso:2023trq}; dark nuclei \cite{Wise:2014jva,Wise:2014ola,Hardy:2014mqa,Gresham:2017zqi,Gresham:2017cvl}; mirror sector objects analogous to ordinary stars, white dwarfs, or neutron stars \cite{Kouvaris:2015rea,Giudice:2016zpa,Maselli:2017vfi,Curtin:2019lhm,Curtin:2019ngc,Hippert:2021fch,Gross:2021qgx,Ryan:2022hku,Bai:2023mfi}; or dark-charged primordial black holes \cite{Bai:2019zcd}. Note that charged black holes necessarily have small charge-to-mass ratios for the vector case to avoid a naked singularity, while black holes with scalar hair are not well understood. This makes them less interesting for enhancing the SGWB signal, which we will see requires much larger charge-to-mass ratios. We will treat MDM in a model-independent way in later discussion such that we do not enforce any relation between charge and mass.

\section{Dark force mediation and radiation}\label{sec:EM_emissions}

With the inclusion of the new interaction, the total force between the binary MDM objects takes the form
\begin{equation}
    F = - \frac{G m_1 m_2}{r^2} \left[ 1 - \alpha e^{-\mmed r} (1+\mmed r) \right] \, ,
\end{equation}
where $\alpha$ characterizes the strength of the new forces relative to gravity, with $\alpha = -y^2 q_1 q_2 / (4 \pi G m_1 m_2)$ in a scalar-mediated model like in~\eqref{eq:yukawa} or $\alpha = g^2 q_1 q_2 / (4 \pi G m_1 m_2)$ in a vector-mediated model like in~(\ref{eq:Lvector}).
In principle, $\alpha$ can be either positive or negative, depending on whether the new interaction between the two MDM objects is repulsive or attractive. For binaries to form and merge, $\alpha<1$ is required so the total force is attractive. We will assume an attractive DF with a negative $\alpha<0$ in the remainder of the text.
We will further assume a massless mediator limit of $\mmed\ll r^{-1}$, with which the dark force reduces to an inverse-square law.
We therefore parametrize the total force between the binary as $F=-G^\prime m_1 m_2/r^2$ where $G^\prime=(1-\alpha)G\equiv\GpoG\uu G$ with $\GpoG > 1$.
This leads to a Kepler-like relation
\begin{equation}
\label{eq:kepler}
    \omega^2 = \frac{G' \mtot}{a^3} \, ,
\end{equation}
with $\omega$ as the orbital angular frequency and $a$ as the semimajor axis, and a virial theorem of the total energy
\begin{equation}
\label{eq:energy}
    E = - \frac{G' \mtot^2 \eta}{2 a} \, ,
\end{equation}
where $\mtot=m_1+m_2$ is the total mass of the binary, and $\eta=m_1m_2/m^2$ is the asymmetry between the two masses. In the massive mediator case $\mmed \gg r^{-1}$, $G'$ is replaced by $G$, recovering the usual gravity-only expressions.

In this work, we employ the purely massless mediator limit. This is valid provided the mediator mass is sufficiently small compared to the binary separation distance and the GW energy spectrum. For the scales relevant here, we expect $r \lesssim \bar{x}$ the average separation between MDM objects at matter-radiation equality, which will be defined in Eq.~(\ref{eq:xbar}) and is shorter than of order $0.1\,{\rm pc}$ for the explored model parameter space. This sets an upper bound on the mediator mass of order $6 \times 10^{-23}~\text{eV}$ for solar mass MDM, with a less stringent bound for lighter MDM. Another upper bound comes from the GW frequency sensitivities at various GW observatories. The smallest frequencies are from pulsar timing arrays, which correspond to $f_\text{GW} \sim \text{nHz} \sim 7 \times 10^{-25}~\text{eV}$, though the bound is relaxed if only higher frequency observatories are of interest. If the mediator mass is larger than these bounds, then part of the spectrum must be altered, which is a topic for future work. The mediator mass is also bounded from below as a function of the coupling strength $\alpha$ from, \eg, Bullet Cluster constraints on DM self interaction. For the example of $m_\text{med} \sim \text{nHz} \sim (10~\text{pc})^{-1}$, the bound is $\alpha \lesssim 10^8$ to $10^{10}$, depending on the DM halo profile assumed \cite{Bogorad:2023wzn}. The values of $\alpha$ we consider will easily satisfy all $m_\text{med}$ constraints discussed here.

From the binary's orbital motion we can derive the GW energy spectrum for a single binary from the binary merger process $dE_{\rm GW}/\fGWs$, where the subscript ${,s}$ means the frequency in the source frame (\ie, emitted by the binary). 
The frequency of the GWs emitted from the binary $\fGWs$ is related to the binary orbital frequency $f_{\rm orb}$ by $\fGWs=2\uu f_{\rm orb}=\omega/\pi$.~\footnote{It is well known that binaries with eccentric orbits radiate GWs with all harmonics of the orbital frequency~\cite{Enoki:2006kj}, \ie, $\fGWs=n\uu f_{\rm orb}$. The approach presented here effectively treats all GW emission to be through the $n=2$ harmonic. We have checked the emission through the other harmonics and found their contributions to be negligible. See Appendix~\ref{app:higher_harmonics} for more details on the higher-harmonic emission.}
The GW energy spectrum can then be calculated by the chain rule as 
\begin{align}
\dfrac{dE_{\rm GW}}{d\fGWs}=\dfrac{\dot{E}_{\rm GW}}{\dot{f}_\text{GW,s}}=\dfrac{\pi \dot{E}_{\rm GW}}{\dot{\omega}}=\dfrac{\pi\dot{E}_{\rm GW}}{- \frac{3 \sqrt{2}}{G' \mtot^{5/2} \eta^{3/2}} \sqrt{-E} \dot{E}}\,,
\end{align}
where the dots indicate derivatives with respect to time, and the last equality uses Eqs.~\eqref{eq:kepler} and \eqref{eq:energy} to transfer the time derivative on $\omega$ to that on the binary system's energy $E$.
Note that the $\dot{E}_{\rm GW}$ in the numerator is the energy emission rate through GWs, while $-\dot{E}$ is the emission rate of the binary system's total energy, to which both GW and $\EM$ emission contribute. In other words, $-\dot{E}=\dot{E}_{\rm GW}+\dot{E}_\EM$ (where the sign comes from $\dot{E}<0$ referring to the loss of energy in the binary, whereas $\dot{E}_{\rm GW,\EM}>0$ refer to the energy emitted as radiation).
Specifically, if the $\EM$ emission is negligible then $-\dot{E}\approx \dot{E}_{\rm GW}$, and without any knowledge on the form of $\dot{E}_{\rm GW}$ the well-known spectral form~\cite{Maggiore:08textbook} is obtained:
\begin{align}\label{eq:dEGWdf}
\dfrac{dE_{\rm GW}}{d\fGWs}=\dfrac{\pi^{2/3}}{3} \,G^{\prime 2/3} \,\mtot^{5/3} \,\eta\, \fGWs^{-1/3}\,.
\end{align}
Note that $G^\prime$ is still present in the expression rather than $G$, as the dark interaction is still affecting the orbital motion of the binary.
This spectrum is valid for $G' m/(a_0^3) < \pi^2 f_\text{GW}^2 < G' m/((r_1+r_2)^3)$, where $r_{1,2}$ are the radii of the MDM objects and $a_0$ is the initial semimajor axis of the binary. The spectrum at still higher frequencies depends on the details of the MDM (\eg, black holes have a merger and ringdown period after their event horizons cross \cite{Ajith:2007kx,Zhu:2011bd}). For the sake of model independence, we neglect any higher frequencies emitted when two MDM merge, meaning our SGWB spectra will be underestimated and our limits conservative.

\begin{figure}[t]
    \centering
    \includegraphics[width=0.48\textwidth]{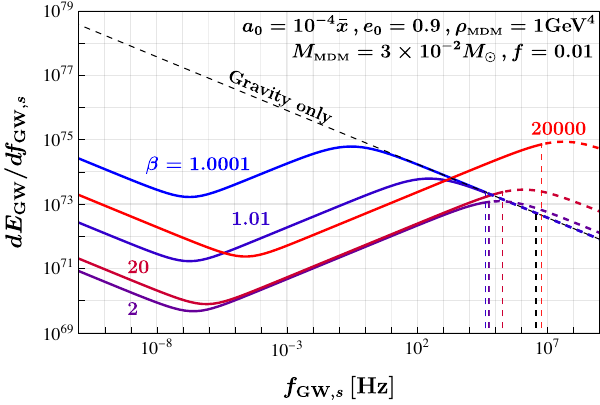} \hspace{2mm}
    \includegraphics[width=0.48\textwidth]{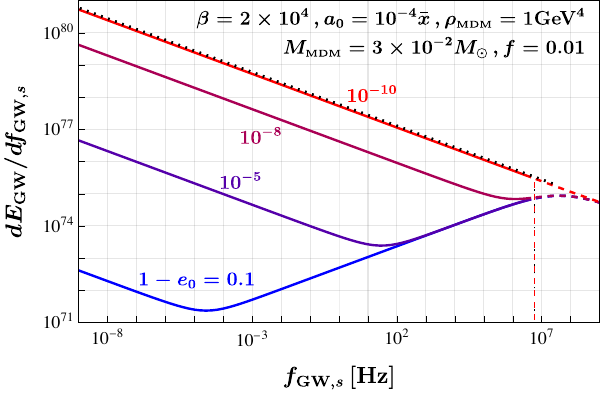}
    \caption{
    {\it Left}: The GW energy spectra  for various $\GpoG$. 
    The rest of the parameters are chosen as $\mdm=3\times 10^{-2}\uu M_\odot$, $e_0=0.9$, $f=0.01$, and $a_0=10^{-4}\bar{x}$ where $\bar{x}$ is defined in Eq.~\eqref{eq:xbar}.
    The colored vertical dashed lines correspond to the frequency cutoff when the MDMs merge assuming $\rhodm=1\uu\GeV^4$, while the dashed curves indicate how the spectrum would continue assuming a still larger $\rhodm$.
    {\it Right}: The GW energy spectra with $\GpoG=2 \times 10^4$ fixed and $e_0$ varied, with all other parameters the same as the left panel. 
    }
    \label{fig:dEGWdfGWs}
\end{figure}

More generically, for an eccentric orbit with orbital eccentricity $e$ and semi-major axis $a$, the GW and $\EM$ energy emission rate $\dot{E}_{\rm GW}$ and $\dot{E}_\EM$ are calculated as
\begin{align}
\langle \dot{E}_\text{GW} \rangle &= \frac{32 G G'^3 \eta^2 \mtot^5}{5 a^5 (1-e^2)^{7/2}} \left(1 + \frac{73}{24}e^2 + \frac{37}{96}e^4 \right) \,,\\
\langle \dot{E}_\EM \rangle &= \frac{G G'^2}{12 \pi} \eta^2 \mtot^4 \left(\frac{g q_1}{\sqrt{G}m_1}- \frac{g q_2}{\sqrt{G}m_2}\right)^2 \frac{1}{a^4} \frac{2+e^2}{(1-e^2)^{5/2}} \, ,
\label{eq:EdotEM}
\end{align}
For concreteness, (\ref{eq:EdotEM}) assumes a vector mediator as the benchmark new force, but the expression for this and other important quantities for a scalar mediator are essentially the same in the massless mediator limit as described in Appendix~\ref{app:emission_rate}.
Note that the evolution of $a$ and $e$ is related. A detailed derivation of the energy emission and the orbital evolution function $a(e)$ is left for Appendix~\ref{app:emission_rate}. 
For simplicity, in the rest of the text we work in the same mass and opposite charge case, $q_1 = -q_2$ and $m_1 = m_2 = \mdm$. In this case, the squared difference of the charge-to-mass ratios in (\ref{eq:EdotEM}) is $16 \pi |\alpha|$.~\footnote{The scalar dark force mediator case is slightly different. Because the force between two MDM objects is always attractive, $q_1$ and $q_2$ are of the same sign, which leads to some mild cancellation for the $\EM$ emission. As discussed around Eq.~\eqref{eq:scalar-charge-mass-ratio}, the charge-to-mass ratio is not fixed for DQNs with different dark quark flavor charges. This cancellation just leads some order-one number for $\langle \dot{E}_\EM \rangle$ of a particular dark binary.} Then with simultaneous GW and $\EM$ emission, the GW energy spectrum from the binary is
\begin{align} \label{eq:dEdfs_complete}
\dfrac{dE_{\rm GW}}{d\fGWs} 
&=\dfrac{\pi\sqrt{a}  \left(37 e^4+292 e^2+96\right) G^{\prime 3/2} \mdm^{5/2}}{3\sqrt{2}\left[10\uu a(1-e^2)(2+e^2)(\GpoG-1)+\left(37 e^4+292 e^2+96\right)G^\prime \mdm\right]}\,,
\end{align}
where $\GpoG \equiv G'/G$, the first term in the denominator comes from $dE_{\rm EM}/dt$, and the second comes from $dE_{\rm GW}/dt$. 
In general, we expect the energy spectrum to have a three-stage broken power-law behavior as shown in the left panel of Fig.~\ref{fig:dEGWdfGWs}, which we briefly analyze below.

The frequency dependence of $dE_{\rm GW}/df_{\rm GW,s}$ comes from the binary orbital semi-major axis $a=(2G^\prime \mdm/(\pi^2 \fGWs^2))^{1/3}$, as well as the dependence of the eccentricity $e$ on $a$ (see, \eg, Eqs.~(\ref{eq:aeGW}) and (\ref{eq:aeEM}) in the Appendix).
When $\fGWs$ is very large, $a$ is small, and the GW emission dominates the EM emission in the denominator of Eq.~(\ref{eq:dEdfs_complete}). Thus, $dE_{\rm GW}/d\fGWs \propto \sqrt{a} \propto \fGWs^{-1/3}$ as in (\ref{eq:dEGWdf}). Similarly, if the initial value of either $a_0$ or $(1-e_0)$ is small, GW emission will dominate throughout the binary inspiral (as shown in the right panel of Fig.~\ref{fig:dEGWdfGWs}).
The GW-dominated regime only appears if $\GpoG$ and/or the MDM radii are sufficiently small for GW emission to dominate below the cutoff frequency; the dashed lines in Fig.~\ref{fig:dEGWdfGWs} indicate how the spectrum would continue in the absence of a cutoff.

In the other extreme when $\fGWs$ is very small, the eccentricity is near its initial value, which in practice is likely to be near unity.
Although the $\EM$ emission dominates the GW emission, $a(1-e^2)$ approaches a constant as suggested by the orbital evolution of the $\EM$-dominated case [see Eq.~\eqref{eq:aeEM} in Appendix~\ref{app:emission_rate} for details], and the denominator of Eq.~\eqref{eq:dEdfs_complete} is insensitive to $\fGWs$. Again the $\fGWs$ dependence comes mainly from the $\sqrt{a}$ in the numerator, and $dE_{\rm GW}/d\fGWs\propto \fGWs^{-1/3}$.

Between the two extremes, the $\EM$ emission dominates the GW emission,  
but the orbital eccentricity $e$ is only mildly dependent on $a$. The energy spectrum then behaves as $dE_{\rm GW}/d\fGWs\propto \sqrt{a}/a\propto \fGWs^{1/3}$.

Note that with increasing $\GpoG \gtrsim 1$, the GW spectrum increases and can exceed the gravity-only GW spectrum (particularly at high frequency). The maximum emission frequency also increases. Both effects are because of the larger orbital frequency from Kepler's law. On the other hand, when $\GpoG \lesssim 1$, increasing $\GpoG$ leads to decreasing GW power, particularly at low frequency. This is because most of the binary's energy is radiated away as $\EM$ emission, but there is very little enhancement to the GW power because the orbital frequency is barely changed. These results refer only to individual binaries. To understand the effect of increasing $\GpoG$ on the SGWB, one must also account for the change in the merger rate, which is the topic of the next section.

\section{Merger Rate and Stochastic GW Spectrum}
\label{sec:rate-SGWB}

Binaries form predominantly in the early universe before matter-radiation equality~\cite{Nakamura:1997sm,Ioka:1998nz,Sasaki:2016jop}, the ensemble of whose coalescence over time forms the SGWB today.
Once the initial distribution of the binaries is given, the merger history can be inferred and the SGWB can be calculated.
The case of binary black hole (BBH) mergers has been intensively studied in previous literature~\cite{Wang:2016ana,Sasaki:2018dmp,Raidal:2017mfl,Atal:2022zux,Braglia:2022icu,Banerjee:2023brn}, where the SGWB abundance spectrum is calculated as
\begin{equation}\label{eq:GW_spec1}
    \Omega_\text{GW} (\fGW) = \frac{\fGW}{\rho_c H_0} \int_0^{z_\text{sup}} dz\frac{R(z)}{(1+z) \sqrt{\Omega_{\rm R} (1+z)^4 + \Omega_{\rm M} (1+z)^3 + \Omega_\Lambda}} \frac{d E_\text{GW}}{d\fGWs}(\fGWs)\Bigg\vert_{\rm BBH} \, , 
\end{equation}
where $R(z)$ is the merger rate at the given redshift $z$, whose upper integration bound $z_\text{sup}$ is discussed below after a change of variables; $H_0$ is the value of the Hubble parameter today; $\rho_c$ is the critical energy density; $\Omega_{{\rm R}, {\rm M}, \Lambda}$ denote the abundance of radiation, matter, and dark energy, respectively; and the GW frequency in the source frame $\fGWs$ is related to the observed GW frequency $\fGW$ via $\fGWs=(1+z)\fGW$.~\footnote{This relationship between $\fGW$ and $\fGWs$, while standard in the literature, is an approximation. It assumes that the GW frequency spectrum is uniformly redshifted with the reference redshift given by the merger redshift. However, smaller frequencies are emitted earlier and actually redshifted more. Nevertheless, this is a reasonable approximation because the time the binary spends at larger frequencies is significantly shorter than the time it spends at smaller frequencies.}
The merger rate $R$ is closely related to the initial distribution of the binaries, which therefore should rely on multiple dynamic variables in principle.
The reason why the integration in~\eqref{eq:GW_spec1} involves only the frequency (which is closely related to the merger lifetime) is that the gravity-only GW energy spectrum of binary MDM coalescence $dE_{\rm GW}/d\fGWs$ depends only on redshift~\cite{Ajith:2007kx,Zhu:2011bd}, and thus all the other variables are can be implicitly marginalized.
For the case where a dark force is involved, on the other hand, the dynamics and the energy spectrum is more complicated as is shown in the previous section, and we therefore will need to keep all the variables.

In the following, we work in the same-mass opposite-charge case. The merger rate $R$ per unit comoving volume 
for a given geometry is related to the binary initial distribution $P$ via
\begin{align} \label{eq:merger-rate-differential}
R(x,y)=\dfrac{1}{2}\dfrac{n_{\MDM}}{2}P=\dfrac{1}{2}\frac{3H^2_0}{8\pi G}\dfrac{f\uu \Omega_{\rm DM}}{2\mdm}P(x,y)\,,
\end{align}
where $n_{\MDM}$ is the MDM comoving number density, $f$ is the fraction of DM contained in MDM, and the factor $1/2$ in the front accounts for the positive and negative charges of the MDM (which is present for a vector mediator but should be omitted for a scalar mediator).
The variables $x$ and $y$ are the initial comoving distances between two nearest-neighboring MDM objects and from the nearest-neighbor center to the next-nearest-neighbor MDM, respectively.
We use the normalized initial distribution $P$ given in~\cite{Ioka:1998nz}, which assumes the MDM objects form randomly in space and takes into account the tidal disruption from the next-nearest-neighbor MDM on the binary evolution. Specifically,
\begin{align}\label{eq:dPdxdy}
P(x,y)\uu dxdy=\dfrac{9 x^2 y^2}{\bar{x}^6} e^{-(y/\bar{x})^3} dx dy\,,
\end{align}
where $\bar{x}$ is taken as the average separation between MDM objects at matter-radiation equality given by
\begin{align}\label{eq:xbar}
\bar{x} = \dfrac{1}{1+z_\text{eq}} \left(\frac{8 \pi G \mdm}{3 H_0^2\uu f\uu \Omega_\text{DM}}\right)^{1/3} \approx 0.1\uu{\rm pc}\left(\dfrac{\mdm}{M_\odot}\right)^{1/3}\left(\dfrac{1}{f}\right)^{1/3}.
\end{align}
In natural units, $(0.1\uu {\rm pc})^{-1} \approx 6 \times 10^{-23} \uu {\rm eV}$, which sets the rough condition for treating the dark force as Coulomb-like in the limit $\mmed \ll \bar{x}^{-1}$, noting that this limit becomes weaker with decreasing $\mdm$.
The semimajor axis $a_0$, semiminor axis $b_0$, and eccentricity $e_0$ of the initial binary orbit can be parametrized as
\begin{align}\label{eq:a_0}
    a_0 &= \dfrac{c_1}{\GpoG} \frac{1}{f} \frac{x^4}{\bar{x}^3} \, ,
    \\
    b_0 &= c_2 \left(\frac{x}{y}\right)^3 a_0 \, ,
    \\ \label{eq:e_0}
    e_0 &= \sqrt{1-\left(\frac{b_0}{a_0}\right)^2}\,,
\end{align}
where $c_1$ and $c_2$ are $\mc{O}(1)$ constants (denoted as $\alpha$ and $\beta$ in~\cite{Ioka:1998nz}). We adopt $c_1=0.4$ and $c_2=0.8$, in alignment with~\cite{Ioka:1998nz}.
A binary decouples from the Hubble flow when the mean energy density from its additional interaction is equal to the drag from the radiation energy density, \ie, $G^\prime \bar{\rho}_\MDM \equiv G^\prime\cdot f \frac{\rho_{\rm eq}}{2} \frac{\bar{x}^3R^3_{\rm eq}}{x^3 R^3} = G \rho_r$, where $R$ is the scalar factor in the FRW metric and the radiation energy density is $\rho_r = \frac{\rho_\text{eq}}{2} \frac{R^4_{\rm eq}}{R^4}$.
This implies the decoupling occurs at $R_{\rm dec}/R_{\rm eq}=(\GpoG f)^{-1} (x/\bar{x})^3<1$. 
The decoupling must be earlier than matter-radiation equality because the average energy density created by a nearest-neighbor MDM pair $\bar{\rho}_\MDM$ redshifts like matter, and thus decreases slower than the cosmic energy density only during radiation domination (RD). 
As a result, any MDM pair that has not already decoupled during RD will not decouple during matter domination (MD)~\cite{Nakamura:1997sm,Ioka:1998nz,Sasaki:2016jop}.
Further refinements to this calculation which include the effects of clustering and binary disruptions in both the early and late universe, as well as the case of an extended rather than monotonic mass function, can be found in \cite{Raidal:2017mfl,Raidal:2018bbj,Vaskonen:2019jpv,Hutsi:2020sol}. These effects are neglected in this work because they require numerical N-body simulations to estimate, which must be redone in the presence of a DF and is thus beyond the scope of this work. Note that binaries that form at later times are generally expected to have a much smaller merger rate than those that form in the early universe and can therefore be neglected \cite{Ali-Haimoud:2017rtz}.

In practice, it is more convenient to use the initial orbital eccentricity $e_0$ and the lifetime $\tau$ of the binaries to express and evaluate the relevant quantities instead of using $x$ and $y$.
In terms of these variables [using (\ref{eq:tau_EM}) as an approximation for the merger lifetime assuming the emission of the dark force carrier dominates the gravitational emission],
\begin{align}\label{eq:a0_tau}
a_0=&\left(\dfrac{4 G^2(\GpoG-1)\GpoG\mdm^2\tau}{h(e_0)}\right)^{1/3}=\left(\dfrac{\tau/\bar{\tau}}{h(e_0)}\right)^{1/3}\bar{x}\,,
\end{align}
where 
\begin{align} \label{eq:taubar}
\bar{\tau}&=\dfrac{2\pi}{3f\uu G(\GpoG-1)\GpoG H^2_0\mdm(1+z_{\rm eq})^3\Omega_{\rm DM}}\,,\\
h(e_0)&=e^{-4}_0(1-e_0^2)^{5/2} \left(1-\sqrt{1-e^2_0}\right)^2\,.
\end{align}
To avoid possible confusion, let $\mc{P}(\tau,e_0)$ denote the normalized distribution function after proper change of variable, including the Jacobian for changing the integration variables from those in (\ref{eq:dPdxdy}).
The SGWB is thus calculated as
\begin{align}\label{eq:GW_spec2}
\Omega_{\rm GW}(\fGW)=\dfrac{\fGW}{\rho_c}\int^{e_{0,{\rm max}}}_{\sqrt{1-c^2_2}} de_0\int d\tau \dfrac{n_{\MDM}}{4}\mc{P}(\tau,e_0)\dfrac{dE_{\rm GW}}{d\fGWs}\left[(1+z(t))\fGW\right]\,,
\end{align}
where $dE_{\rm GW}/d\fGWs$ is the GW energy spectrum Eq.~(\ref{eq:dEdfs_complete}), $t(\tau, e_0)=t_{\rm dec}(\tau,e_0)+\tau$ is the time when the binary merges, and $t_{\rm dec}$ is the cosmic time when a binary with a given initial condition decouples from the Hubble flow.
The discussion below Eq.~\eqref{eq:e_0} implies the binary decoupling happens at redshift
\begin{align}\label{eq:z_dec}
1+z_{\rm dec}=\left(\dfrac{2\pi\uu c^3_1(1+z_{\rm eq})}{3H^2_0 G \mdm(\GpoG-1)\Omega_{\rm DM}}\dfrac{h(e_0)}{\tau}\right)^{1/4}\,.
\end{align}
Apparently $z_{\rm dec}$ $(t_{\rm dec})$ is a monotonously decreasing (increasing) function on both $e_0$ and $\tau$, and thus $t=t_{\rm dec}+\tau$ monotonously increases with respect to $\tau$, enabling a one-to-one correspondence between the two quantities.

With Eq.~\eqref{eq:GW_spec2} given, we need to find the integration range of $\tau$ and $e_0$.
The integration range of $\tau$ is subject to three different constraints which are imposed not only directly on $\tau$ but also indirectly on $t$. The first constraint arises from the cosmological requirement that the binary merges after MDM formation and before today.
For the representative model in Sec.~\ref{sec:model} and similar models, we may generically expect the formation of MDM happens at a cosmic temperature of $T_{\rm form}\sim \rhodm^{1/4}$~\cite{Bai:2018dxf,Bai:2022kxq}, or in other words, at a cosmic redshift of $1+z_{\rm form}\sim \rhodm^{1/4}/T_{\gamma,0}$. The constraint thus requires $t_{\rm form}<t <t_0$. \footnote{Binaries that have not yet merged would also contribute to the stochastic GW background.}

The second constraint arises from the requirement that the binary system decouples from the Hubble flow before matter-radiation equality, which sets $x<(f\,\GpoG)^{1/3}\bar{x}$.
The fact that $x=(a_0 f\uu \GpoG \bar{x}^3/c_1)^{1/4}$ together with Eq.~\eqref{eq:a0_tau} gives an upper bound on $\tau$ as
\begin{align}
\tau<c^3_1 \,f  \,\GpoG \, h(e_0)\,\bar{\tau}\,.
\label{eq:t_max_2}
\end{align}

The third constraint comes from the frequency band of an individual merger event. As explained below (\ref{eq:dEGWdf}),
\begin{align}
\frac{G^\prime \mtot}{a^3_0}<\pi^2 \left[1+z(t)\right]^2 \fGW^2< \frac{G^\prime \mtot}{(2\rdm)^3}\,.\label{eq:tau_constraint_3}
\end{align}
As redshift $z(t)$ is a monotonically decreasing function of $t$, the right-hand (RH) inequality implies a lower bound of $t$.
The left-hand (LH) inequality is more tricky due to the $\tau$-dependence of $a_0$, as given in Eq.~\eqref{eq:a0_tau}.
Collecting all the $t$- and $\tau$-dependent terms, the LH inequality is translated to
\begin{align}
\Longrightarrow \; &
\left[1+z(t)\right]^2 \tau>\frac{1}{\pi^2 \fGW^2} 2 G' \mdm \bar{x}^{-3} \bar{\tau}\uu h(e_0) \,.\label{eq:t_max_3}
\end{align}
For any given $e_0$ the right-hand side (RHS) of~\eqref{eq:t_max_3} evaluates to a constant, while the left-hand side (LHS) could be rather complicated.
If $t$ is deep inside the RD regime of the cosmic history (\ie, $t$ and $\tau$ are rather small), then we expect $t_{\rm dec}=\sqrt{\tau}/(2H_0\sqrt{\Omega_{\rm R}}\mc{C}^2)$ where $\mc{C}$ is defined through Eq.~\eqref{eq:z_dec} as $1+z_{\rm dec}=\mc{C}/\tau^{1/4}$.
Then the LHS evaluates to
\begin{align}
\left[1+z(t)\right]^2 \tau=\dfrac{\tau}{2H_0\sqrt{\Omega_{\rm R}}\left(\tau+\frac{\sqrt{\tau}}{2H_0\sqrt{\Omega_{\rm R}}\mc{C}^2}\right)}\,,
\end{align}
which is an increasing function of $\tau$.
Meanwhile, for larger $\tau$, we generically have $t\approx\tau$ and the LHS reduces to the monotonous decreasing function $[1+z(\tau)]^2 \tau$.
In other words, the LH inequality of Eq.~\eqref{eq:tau_constraint_3} may impose not only an upper bound on $t$ (or equivalently, $\tau$) but also a lower bound as well.

For the range of the $e_0$ integration, the lower bound is determined by the requirement $y>x$~\cite{Ioka:1998nz}, \ie, by definition the next-nearest-neighbor MDM object should be further away. With $y=x\uu c^{1/3}_2 (1-e^2_0)^{-1/6}$ this requirement is translated to $e_0>\sqrt{1-c^2_2}$.
The upper bound on $e_0$, $e_{0,{\rm max}}$, is determined by the phase space of the $\tau$-integration.
Note that it is not guaranteed that there is always a valid phase space for the $\tau$ integration under the three constraints described above. In fact, the allowed range of $\tau$ decreases as $e_0\to 1$ and eventually vanishes completely, which thus determines $e_{0,{\rm max}}$ in Eq.~\eqref{eq:GW_spec2}.

\section{Properties of and constraints on the SGWB}\label{sec:results}
\subsection{An analytic understanding of the SGWB spectrum}\label{sec:SGWB_shape}

\begin{figure}[t]
    \centering
    \includegraphics[width=0.6\textwidth]{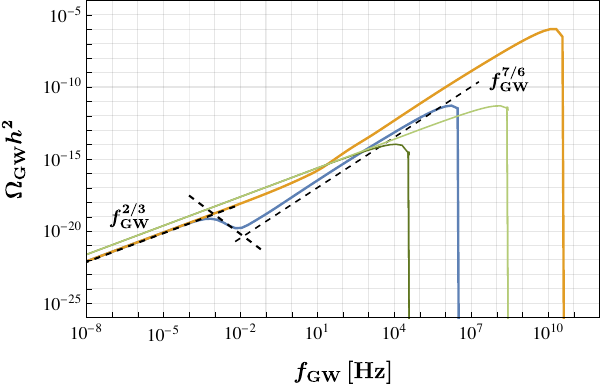}
    \caption{An illustration of the shape of the SGWB from dark binaries. For concreteness, the two curves are calculated with $\mdm=10^{-7} M_\odot$, $f=0.01$, $\GpoG=10^4$, and $\rhodm=1\uu{\rm GeV}^4$ (blue, lower) or $(100\uu \GeV)^4$ (orange, upper), respectively. Also overlaid as dashed lines are the analytic estimates for the blue curve's  power-law behavior given in Eqs.~\eqref{eq:OmegaGW_low}, \eqref{eq:OmegaGW_high}, and (\ref{eq:OmegaGW_middle}). 
    For comparison, the solid green curves show the spectrum for the gravity-only case, where the darker/lighter green curves share the same $\mdm$, $f$, and $\rhodm$ as the blue/orange curves, respectively. The dark binaries have a larger $\fGW$ cutoff when having an attractive DF between them, as the the orbital frequency is increased by the DF.
    }
    \label{fig:spectra_shape}
\end{figure}

Due to the complexity of the energy spectrum at the source $dE_{\rm GW}/d\fGWs$, the SGWB behaves as a two- or three-stage broken power-law following the energy spectrum behavior in \eqref{fig:dEGWdfGWs}.
See Fig.~\ref{fig:spectra_shape} for an illustration, which plots numerical calculations and analytic estimates for $\Omega_\text{GW} h^2 (f_\text{GW})$ with $h=H_0/(100 \, \text{km/s/Mpc})=0.678$~\cite{Planck:2018vyg}.
Below we will briefly discuss the shape of $\Omega_{\rm GW}$ at low and high frequencies, and leave the detailed analytic calculations (including the middle frequencies) for Appendix~\ref{app:OmegaGW_shape}.

In the low $\fGW$ region, the SGWB is dominated by mergers that largely happen when the universe is radiation dominated, and the orbital eccentricity is largely not softened from one. 
This is because most binaries merge with very short lifetimes, and all binaries with sufficiently large initial semimajor axis contribute to the low-frequency SGWB (as opposed to the high-frequency background, which due to redshift comes exclusively from more recent mergers).
In this regime $t_{\rm dec}\gg \tau$, \ie, the binaries merge swiftly after their decoupling. For a given merger to contribute a particular GW frequency $f_\text{GW}$ to the GW spectrum today, it must merge sufficiently late so that the maximum frequency at the source is not redshifted away. From the RH inequality in (\ref{eq:tau_constraint_3}) and the expression in (\ref{eq:z_dec}) with $t \approx t_\text{dec}$, there is a corresponding lower bound on $\tau$ for the integration in (\ref{eq:GW_spec2}) given by
\begin{align}
\label{eq:taumin}
\tau_{\rm min}\approx\left(\dfrac{3\pi \fGW^2}{G^\prime \rhodm}\right)^2\dfrac{2\pi c^3_1(1+z_{\rm eq})h(e_0)}{3H^2_0 G \mdm(\GpoG-1)\Omega_{\rm DM}}\,.
\end{align}
Also, in this frequency regime there will be a competition between the contribution to the spectrum from the GW radiation and the $\EM$ radiation at different values of $e_0$.
At relatively small $e_0$ the $\EM$ emission is more efficient, while above a certain threshold $e_0>e_{0,{\rm th}}$ the GW emission dominates. The latter turns out to make up the majority of the spectrum such that effectively we only need to perform the $e_0$-integration on the interval, $[e_{0,{\rm th}},1)$.
The determination of $e_{0,{\rm th}}$ is discussed in Appendix.~\ref{app:OmegaGW_shape}.
Both the $\tau$- and the $e_0$-integration can be done analytically after these approximations, evaluating to a spectrum of 
\begin{align}\label{eq:OmegaGW_low}
\Omega_{\rm GW}\approx  
\left( \dfrac{85^{6} \pi^{8} \Gamma^{9}(\frac{7}{3}) f^{18} G^{14}\GpoG^{17}H^4_0\mdm^{14}(1+z_{\rm eq})\Omega^{17}_{\rm DM}}{2^{77} 3^{4} c^{12}_1 c^{12}_2 (\GpoG-1)^{6}} \right)^{\frac{1}{15}} f^{2/3}_{\rm GW}\,.
\end{align}
Note that this result is independent of $\rhodm$, which matches the numeric result in Fig.~\ref{fig:spectra_shape}. This is as expected: low-frequency GWs should not depend on $\rhodm$, which sets the high-frequency cutoff in the binary source frame in $dE_\text{GW}/df_\text{GW,s}$.

In the high frequency regime close to the spectrum cutoff the dominant contribution changes.
This regime primarily receives contributions to the SGWB from binaries that satisfy $\tau\gg t_{\rm dec}$ such that $t\approx\tau$ with binary coalescence occurring during MD. As a result, the orbital eccentricity has had time to soften, and to reasonable approximation $e=0$. The $\EM$ emission dominates the GW emission, leading to
\begin{align}
&1+z(t)\approx\left(\dfrac{2}{3H_0\sqrt{\Omega_{\rm M}}\tau}\right)^{2/3}\,,\\
&\dfrac{dE}{d\fGWs}\approx \left(\dfrac{2}{3H_0\sqrt{\Omega_{\rm M}}\tau}\right)^{2/9} \dfrac{4\times 2^{1/3}\pi^{4/3}G^{\prime\uu 4/3}\mdm^{7/3} \fGW^{1/3}}{5(\GpoG-1)}\,.
\end{align}
The inner $\tau$-integration in Eq.~\eqref{eq:GW_spec2} can then be evaluated analytically, and the outer $e_0$-integration can be done with a saddle-point approximation, which gives 
\begin{align}\label{eq:OmegaGW_high}
\Omega_{\rm GW}\approx \dfrac{4\times 2^{\frac{5}{6}}\, 13^{\frac{1}{18}} \pi^{\frac{59}{36}} f^{\frac{13}{9}} G^{\frac{55}{36}}\GpoG^{\frac{67}{36}}(1+z_{\rm eq})^{\frac{1}{3}}\rhodm^{\frac{1}{12}} H^{\frac{1}{9}}_0 \mdm^{\frac{13}{9}} \Omega^{\frac{10}{9}}_{\rm DM}}{45\times 3^{7/36} e^{13/9} c^{1/3}_1 c^{5/9}_2 (\GpoG-1)^{8/9} \Omega^{1/18}_{\rm M}}\fGW^{7/6}\,.
\end{align}
This also provides a good match to the numeric results as illustrated in Fig.~\ref{fig:spectra_shape}. Note that the power-law index of $7/6$ depends on our underlying MD assumption. (Assuming RD changes the index to $11/9$ with minimal change to the amplitude, which we discuss in Appendix~\ref{app:OmegaGW_shape}.) 
Noticing that the frequency cut-off in the spectrum is determined by maximum frequency of the binary system merging at present, \ie, the upper limit in Eq.~\eqref{eq:tau_constraint_3} taking $z=0$, the peak frequency and amplitude of the SGWB can be inferred from (\ref{eq:OmegaGW_high}) as 
\begin{align}\label{eq:fGW_max}
f_{\rm GW,max}=&\dfrac{1}{\pi}\left[\dfrac{2G^\prime \mdm}{(2\rdm)^3}\right]^{1/2}=\left(\dfrac{G^\prime \rhodm}{3\pi}\right)^{1/2}=(1 \times 10^6 \,{\rm Hz})\left(\dfrac{\GpoG}{10^3}\right)^{1/2}\left(\dfrac{\rhodm}{1\,\GeV^4}\right)^{1/2}\,,\\ \vspace{3mm}
\label{eq:OmegaGW_max}
\Omega_{\rm GW,peak}\approx &
\dfrac{2^{\frac{9}{4}}\, 13^{\frac{1}{18}} \pi^{\frac{19}{18}}\rhodm^{\frac{2}{3}}}{45\times 3^{\frac{7}{9}} e^{\frac{13}{9}}\Omega^{\frac{1}{18}}_{\rm M}}\left(\dfrac{1+z_{\rm eq}}{c_1}\right)^{\frac{1}{3}}\left(\dfrac{f^{13}G^{19}\GpoG^{22}H_0\mdm^{13}\Omega^{10}_{\rm DM}}{c^5_2(\GpoG-1)^8}\right)^{\frac{1}{9}}\,.
\end{align}

For comparison, the SGWB signal from dark binaries that only couple to gravity without a DF is shown as the green curves in Fig.~\ref{fig:spectra_shape}. 
They are calculated using Eq.~\eqref{eq:GW_spec1}, with the merger rate $R(z)$ derived from~\cite{Ioka:1998nz}.
For some cases, the DF emission can lead to either an enhancement or suppression of the SGWB compared to the gravity-only case. Whether an enchancement or suppression (or both at different frequencies, as for the blue curve) occurs depends on $\rhodm$ as shown in Fig.~\ref{fig:spectra_shape} as well as $\GpoG$.

\begin{figure}[t]
    \centering
    \includegraphics[width=0.48\textwidth]{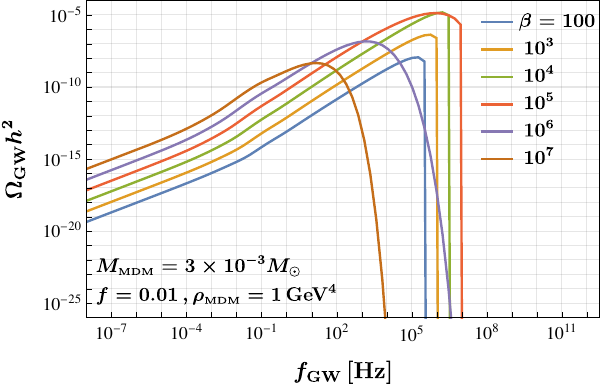} \hspace{2mm}
    \includegraphics[width=0.48\textwidth]{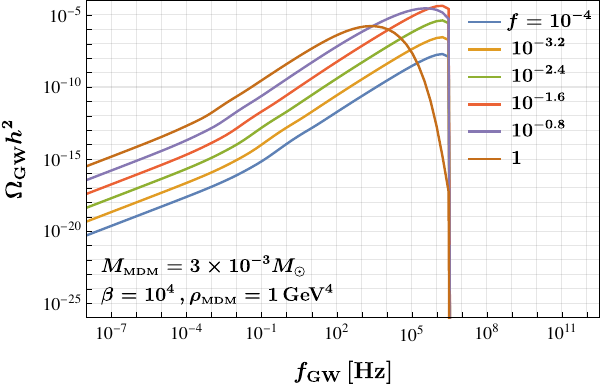}
    \caption{{\it Left:} The calculated GW spectra for $f=0.01$ and various $\GpoG$. {\it Right:} The calculated GW spectra for $\GpoG=10^4$ and various $f$. The rest of the parameters are fixed to be $\mdm=3\times 10^{-3}M_\odot$ and $\rhodm=1\uu\GeV^4$. As $\GpoG$ and $f$ increase beyond the limit in \eqref{eq:constraint-no-high-suppression}, the amplitude at the predicted peak is severely suppressed and the actual peak is shifted to the lower frequency. 
    }
    \label{fig:GpoG_f_upperlim}
\end{figure}

There is a subtlety in the calculation of $\Omega_{\rm GW}$ near $\Omega_{\rm GW,peak}$.
It can be seen from Eq.~\eqref{eq:dPdxdy} that the probability distribution $\mc{P}$ in Eq.~\eqref{eq:GW_spec2} involves an exponential function, which is explicitly written as 
\begin{align}
\mc{P}\propto\exp\left[-\left(\dfrac{3c^4_2 f^4 \GpoG^4 (\GpoG-1)H^2_0 G\uu \mdm (1+z_{\rm eq})^3\Omega_{\rm DM} \tau}{2\pi c^3_1(1-e^2_0)^2h(e_0)}\right)^{1/4}\right]\,.
\end{align}
This exponential function implies that the integration in~\eqref{eq:GW_spec2} is strongly influenced by the parameter space where $\tau\sim\tau_{\rm min}$.
If the expression in the exponential is much smaller than $-1$ even when $\tau\sim\tau_{\rm min}$, the contribution to $\Omega_{\rm GW}$ from the corresponding parameter space should be highly suppressed. In fact, this is related to the validity of the saddle-point approximation we made in the $\tau$-integration in obtaining Eq.~\eqref{eq:OmegaGW_high}.
Requiring that the expression in the exponential be greater than negative one so that the high frequency spectrum is not suppressed sets a requirement
\begin{align}
\label{eq:constraint-no-high-suppression}
\GpoG \left(\dfrac{f}{0.01}\right)^{4/5}\left(\dfrac{\mdm}{10^{-10}M_\odot}\right)^{1/5}\left(\dfrac{\tau_{\rm min}}{t_{\rm now}}\right)^{1/5} \lesssim 8\times 10^6\, \;\quad  \text{(for no high-frequency suppression)},
\end{align}
for $c_1=0.4$ and $c_2=0.8$.
In Fig.~\ref{fig:GpoG_f_upperlim} we show examples of such suppression when $\GpoG$ and $f$ are large.
Focusing on $\GpoG$, if $\GpoG$ is very large, a vast majority of binaries will merge very early, shortly after forming, and only an exponentially small portion with large initial separations will merge later. The binaries that merge early have their GW signals redshifted, leading to a suppression of the peak frequency. And the larger $\GpoG$ becomes, the earlier the binaries merge, leading to larger redshift factors and $\Omega_\text{GW}$ peaking at still smaller frequency. Mathematically, this is captured by the $f_\text{GW}$ dependence in $\tau_\text{min}$. 

\subsection{Experimental sensitivity}
\subsubsection{Ground- and satellite-based interferometers}

The possibility for a SGWB signal $\Omega_{\rm GW}(\fGW)$ to be visible by an experiment is characterized by the signal-to-noise ratio (SNR) $\varrho$. If $\varrho$ is larger than a threshold value $\varrho_{\rm th}$ then the SGWB can be considered as detectable by the experiment. The sensitivity of the experiment is quantified by the detector noise energy density spectrum $\Omega_{\rm noise}(\fGW)$. If the experiment consists of a single detector, the SNR~\footnote{In some works it is $\varrho^2$ instead of $\varrho$ that is defined with the right hand side of Eq.~\eqref{eq:snr_single_detector}. In that case the SNR is defined with respect to the GW amplitude rather than energy density, see discussions in, \eg,~\cite{Maggiore:08textbook}. We use $\varrho$ to remain aligned with the SNR definition~\eqref{eq:snr_corr} from multiple-detector cross-correlation.} is calculated by~\cite{Moore:2014lga}
\begin{align}\label{eq:snr_single_detector}
\varrho=\int d\fGW\dfrac{\lvert h_c(\fGW)\rvert^2}{\fGW^2 S_n}=\int d\fGW \dfrac{3H^2_0\uu \Omega_{\rm GW}}{2\pi^2 \fGW^4 S_n} \, ,
\end{align}
where $h_c$ is dimensionless GW signal strain and $S_n$ is strain noise auto power spectrum of the detector, which is related to the noise energy density spectrum spectrum via $\Omega_{\rm noise}=2\pi^2 \fGW^3 S_n/(3H^2_0)$.
If the experiment contains multiple detectors with uncorrelated noise, the use of the matched filtering technique can greatly improve the sensitivity in a search for a correlated SGWB signal. Then, the SNR is instead calculated as~\cite{Schmitz:2020syl}
\begin{align}\label{eq:snr_corr}
\varrho^2=n_{\rm det}T_{\rm obs}\int d\fGW\left(\dfrac{\Omega_{\rm GW}}{\Omega_{\rm noise}}\right)^2\,,
\end{align}
where $T_{\rm obs}$ is the detector's observation time, 
and $n_{\rm det}=2$ for cross-correlation between signals from different detectors. 
Note that one can repeat the match filtering procedure of cross-correlation, but take all the signals to be from the same detector, which is known as auto-correlation (self-correlation).
Applying the match filtering technique in the case of auto-correlation requires either perfect instantaneous noise monitoring or that the noise strain of the detector not be correlated with itself.
If this can be achieved then Eq.~\eqref{eq:snr_corr} can be used for a single detector, with $n_{\rm det}$ set to 1.

For the existing and proposed experiments, we mainly consider SKA~\cite{Carilli:2004nx}, LISA~\cite{LISA:2017pwj}, BBO~\cite{Corbin:2005ny}, and the 
aLIGO-aVirgo network (HLV, HL for the Hanford and Livingston detectors of aLIGO) \cite{LIGOScientific:2014pky,VIRGO:2014yos} as the representatives at different frequency bands, among which the cross-correlation technique is applicable on SKA, BBO, and HLV.
LISA is a three-satellite Michelson interferometer, and thus is a single detector. We therefore will focus on the SNR calculated from Eq.~\eqref{eq:snr_single_detector}, while we also present results based on an ideal auto-correlation on LISA. Taiji~\cite{Hu:2017mde} has a configuration close to LISA and a similar noise power spectrum. If it is possible to cross correlate the signals at the two experiments~\cite{Ruan:2020smc}, a much better sensitivity can be achieved with matched filtering for GW signals at the mHz frequency range.
The observation time $T_\text{obs}$ is taken as one year for all experiments except SKA, which is taken as 20 years.
The threshold SNR is taken to be $\varrho_{\rm th}=1$, though the results are not very sensitive to this choice (it is also equivalent to a longer observation time with a higher SNR threshold $\varrho_\text{th}=\sqrt{T_\text{obs}/(1 \, \text{yr})}$, where 1 year would be replaced by 20 years for the case of SKA).

\begin{figure}[t!]
    \centering
    \includegraphics[width=0.6\textwidth]{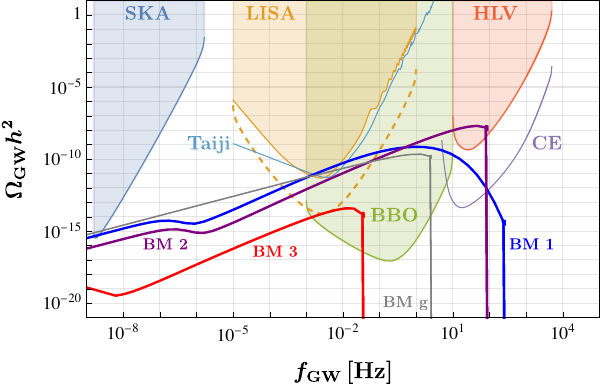}
    \caption{Comparisons between some calculated benchmark SGWB from dark binaries with the PLIS curves of SKA, LISA (dashed), BBO, CE, and HLV and with the noise energy density spectrum $\Omega_{\rm noise}$ of LISA (solid) and Taiji. The PLIS of LISA is presented dashed due to the use of auto-correlation. The sensitivity curves are taken from Ref.~\cite{Schmitz:2020syl}. Note that the sensitivity curves should only be used as a rough reference of $\varrho_{\rm th}=1$ rather than a strict criteria, as discussed in the main text.
    The model parameters of the three benchmarks BM 1-3 are $(\GpoG,\mdm,f,\rhodm^{1/4})=(10^4,0.1\uu M_\odot,0.25, 10\uu{\rm MeV})$, $(10^3,0.1\uu M_\odot,0.25, 10\uu{\rm MeV})$, and $(10^4,0.1\uu M_\odot,0.01, 100\uu{\rm keV})$, respectively, while the gravity-only ``BM g'' has parameters $(1,0.1\uu M_\odot,0.25, 10\uu{\rm MeV})$ chosen to match those of BM 1 and 2.} 
    \label{fig:low_freq_signal}
\end{figure}

In Fig.~\ref{fig:low_freq_signal} we show the comparisons between some calculated benchmark (BM) SGWB with the power-law integrated sensitivity (PLIS) $\Omega_{\rm PLIS}h^2$ of SKA, LISA, BBO, CE \cite{Reitze:2019iox}, and HLV, along with the noise energy density spectra for LISA and Taiji. Note that BM 1, 2, and g are presently excluded by microlensing constraints \cite{Macho:2000nvd,EROS-2:2006ryy,Wyrzykowski_2011,Croon:2020wpr,Bai:2020jfm} and are presented for illustration only.
The PLIS curves and noise energy density spectra are taken from Ref.~\cite{Schmitz:2020syl}.
Note that these PLIS curves are obtained by assuming the GW signal spectra take the form of an exact power-law, and thus should be used only as references for the experiments' sensitivities with $\varrho_{\rm th}=1$. 
To correctly account for the experiments' sensitivities on the parameter space, one still needs to use Eq.~\eqref{eq:snr_single_detector} or Eq.~\eqref{eq:snr_corr} to calculate the SNR.

\begin{figure}[t!]
    \centering
    \includegraphics[width=0.85\textwidth]{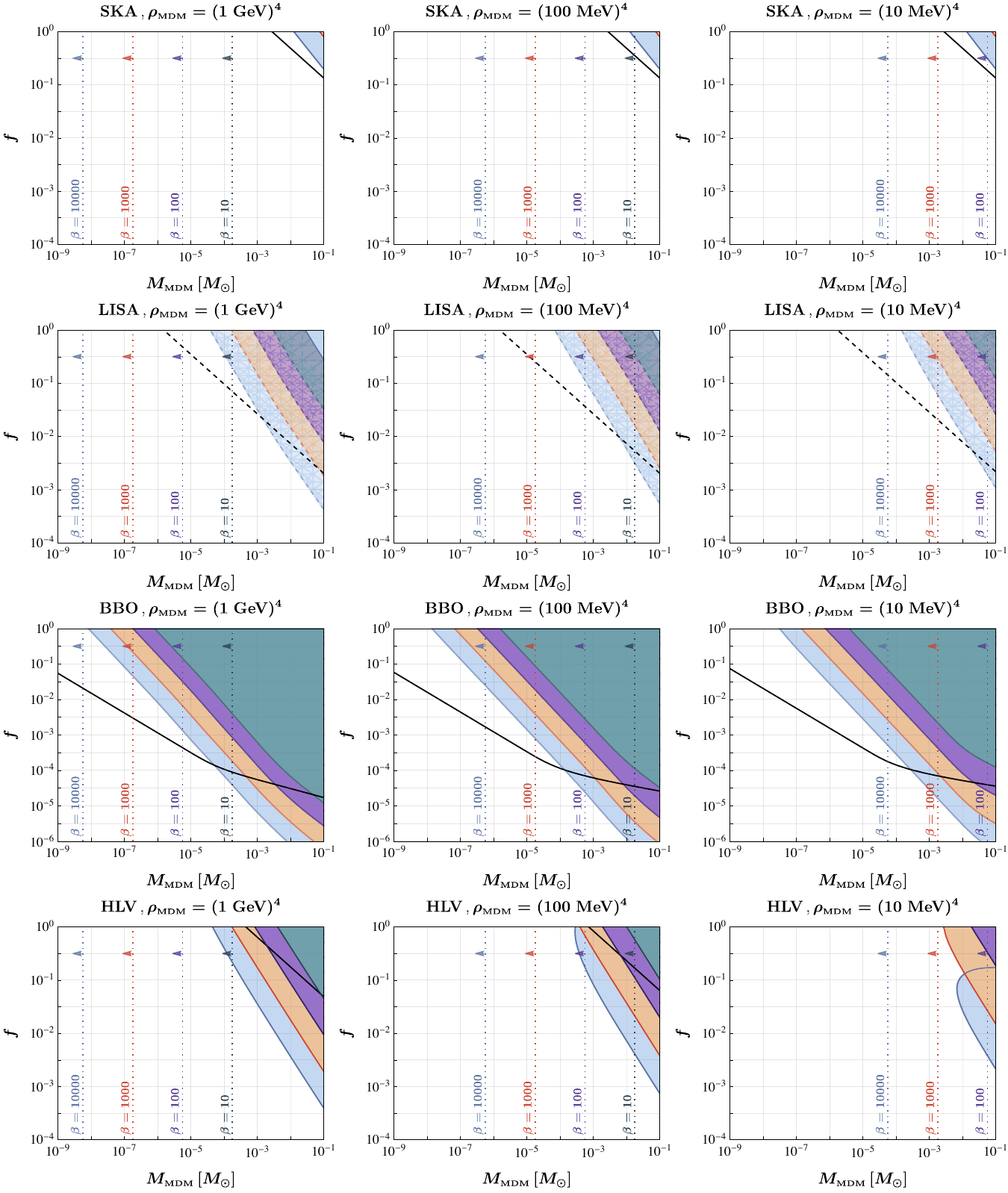}
    \caption{
    Contours of $\varrho=1$ of the SGWB from dark binaries on SKA, LISA, BBO, and HLV for various $\GpoG$ and MDM densities $\rhodm$. The blue/orange/purple/green contours (generally ordered from left to right) are for $\GpoG=10000, 1000, 100$, and $10$, respectively, and the values of $\rhodm$ used are shown above the panels. Black contours refer to the gravity-only case. 
    For LISA (second row), the dashed lines assume a perfect auto-correlation, while the solid lines assume auto-correlation is not applicable.
    The vertical dashed lines are upper bounds on $\mdm$ for given $\GpoG$ assuming the most naive dark quark nugget model, as estimated in Eq.~\eqref{eq:alpha_finite_density}.
    {\it Note the plot range on the vertical axis is larger for BBO (third row) than for the other observatories.}
    }
    \label{fig:SNR_contours}
\end{figure}

The contours of $\varrho=1$ for SKA, LISA, BBO, and HLV for various $\GpoG$ and $\rhodm$ are shown in Fig.~\ref{fig:SNR_contours}, where in the calculation we use the detector $\Omega_{\rm noise}$ provided in~\cite{Schmitz:2020syl}. 
On the top row, the SKA constraints do not change much across different panels because  the low frequency spectrum~\eqref{eq:OmegaGW_low} is insensitive to $\rhodm$.
Given the spectral shape, BBO turns out to be the most powerful experiment to search for or constrain the SGWB from the dark binaries.
The power-law spectral shape largely implies the contours of constant $\varrho$ to be straight lines, while at large $f$ and $\GpoG$ the contours may start to become curved, where the interferometers start to lose sensitivities on the SGWB, as seen in the bottom panels.
This is exactly the result of the high-frequency suppression discussed at the end of the previous subsection and illustrated in both Fig.~\ref{fig:GpoG_f_upperlim} and BM 1 and 2 of Fig.~\ref{fig:low_freq_signal}. 
These two BMs illustrate why in the bottom-right panel of Fig.~\ref{fig:SNR_contours} HLV is sensitive to MDM up to $f=1$ for $\GpoG=10^3$ but not for $\GpoG=10^4$---the larger $\GpoG$ of BM 1 leads to a lower frequency cutoff compared to BM 2 in the SGWB spectrum.
Also shown in Fig.~\ref{fig:SNR_contours} are the upper bounds on the mass of the MDM assuming the naive dark quark nugget model estimated in Eq.~\eqref{eq:alpha_finite_density} with $\beta = 1 - \alpha > 1$.
This bound can be more stringent than the interferometer constraints, especially when $\rhodm$ is large, but it is model dependent and therefore less universal than the constraints from the interferometer.

It is clear from Figs.~\ref{fig:low_freq_signal} and \ref{fig:SNR_contours} that BBO has the potential to search the widest possible parameter space of all GW interferometers considered here.  Clearly, any signal observable by HLV should also be observable by BBO, given the more gradual slope of the $\Omega_\text{GW}$ signal compared to the experimental sensitives. By the same logic, if a signal is observable at SKA or LISA and has a sufficiently large cutoff frequency, it will also be observed by BBO.
In fact, any signal that is visible to any of the other observatories must also be observable by BBO. 
This may seem counterintuitive---could not a signal be large enough to be visible at LISA or SKA, but then have a small enough cutoff frequency to be invisible to BBO?
Indeed, Eqs.~\eqref{eq:fGW_max} and~\eqref{eq:OmegaGW_max} suggest that the peak frequency scales like $\GpoG^{1/2}\rhodm^{1/2}$, and the peak amplitude scales like $\GpoG^{14/9}\rhodm^{2/3}$, so naively it seems like such a possibility can be realized by increasing $\GpoG$ and decreasing $\rhodm$.
For example, in BM 3 in Fig.~\ref{fig:low_freq_signal}, the parameters $\rhodm^{1/4}=100\uu{\rm keV}$ and $\GpoG=10^4$ result in a peak frequency of the spectrum between $10^{-1}$--$10^{-2}$Hz, and an amplitude of $\Omega_{\rm GW}h^2<10^{-13}$.
A naive calculation suggests that to shift the peak further down by about two orders of magnitude while increasing the amplitude by $\sim 10$ such that the signal is visible only to LISA auto-correlation we will need $\GpoG\sim 1 \times 10^8$ and $\rhodm \sim (1\uu{\rm keV})^4$. 
However, numeric calculations show that with such a large $\GpoG$ the peak of the spectrum receives a heavy suppression compared with the naive scaling, as suggested in the end of Sec.~\ref{sec:SGWB_shape}.

\subsubsection{High frequency GW signals}
In additional to the searches on ground- and satellite-based interferometers, the SGWB is also subject to other constraints from direct searches or indirect observations.
In particular, the GWs contribute to the cosmic energy budget as extra radiation degrees of freedom and therefore are constrained by the cosmic microwave background (CMB).
The constraint is conventionally parametrized in terms of the number of extra effective neutrino species $\Delta N_{\rm eff}$ through
\begin{align}
\Delta\rho_{\rm rad}=\dfrac{\pi^2}{30}\dfrac{7}{4}\left(\dfrac{4}{11}\right)^{4/3}\Delta N_{\rm eff}\,T^4\,.
\end{align}
The current bounds from CMB measurements are $\Delta N_{\rm eff}\lesssim 0.3$~\cite{Planck:2018vyg}, and the future CMB-Stage 4 experiment can set a bound of $\Delta N_{\rm eff}\lesssim 0.06$ at the 95\% CL~\cite{Abazajian:2019eic}.
However, one should not directly compare this constraint with the $\Omega_{\rm GW}$ spectrum calculated through~\eqref{eq:GW_spec2}, because the $\Delta N_{\rm eff}$ constraint limits the extra radiation energy budget at recombination, while the $\Omega_{\rm GW}$ spectrum from MDM receives a considerable contribution after recombination (similarly, there is no constraint from Big Bang nucleosynthesis).
Therefore, to calculate $\Delta N_{\rm eff}$ from the dark binaries, the upper integration limit on the binary merger time in~\eqref{eq:GW_spec2} must not exceed the time of recombination.

\begin{figure}[t]
    \centering
    \includegraphics[width=0.48\textwidth]{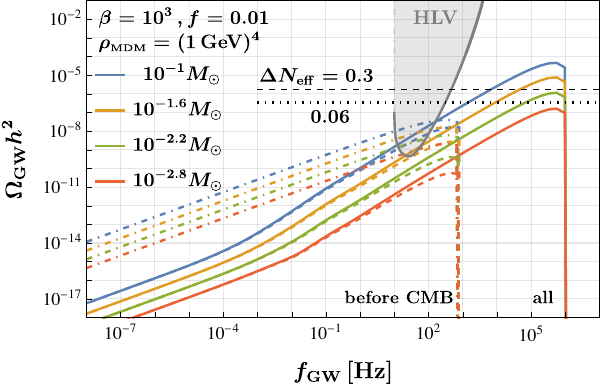}
    \includegraphics[width=0.48\textwidth]{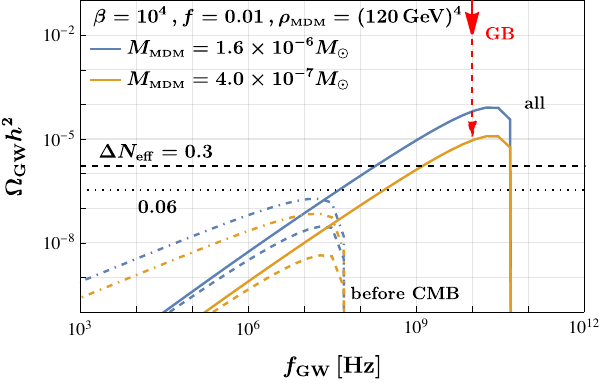}
    \caption{{\it Left}: The calculated SGWB from contributions only before recombination (dashed) and from all cosmic history (solid), together with the DF radiation before recombination (dot-dashed) for various $\mdm$. For comparison we show the PLIS curve for the aLIGO-aVirgo network as the gray shaded region, as well as $h^2\Delta\rho_{\rm rad}/\rho_{c,0}$ translated from the $\Delta N_{\rm eff}=0.3$ and 0.06 bound from the current and expected future CMB measurements as the horizontal lines. Due to the contribution after the recombination, $\Omega_{\rm GW}$ today can be larger than $\Delta\rho_{\rm rad}/\rho_{c,0}$. {\it Right}: Cases where the SGWB at high frequency can be potentially detected by future Gaussian beam (GB) experiments while remain compatible with the $\Delta N_{\rm eff}$ bound at the recombination. The solid red arrow corresponds to $h_c=10^{-30}$ sensitivity at $\fGW=10\uu{\rm GHz}$ in~\cite{Li:2003tv}, and the dashed red arrow is the projection if the GB strain sensitivity is improved by a factor or 30. The benchmark values are in severe tension with the (model-dependent) upper limits \eqref{eq:alpha_finite_density} and \eqref{eq:alpha-upper-vector} estimated with dark quark nuggets, while the possibility remains interesting for other models.}
    \label{fig:high_freq_signal}
\end{figure}

The left panel of Fig.~\ref{fig:high_freq_signal} shows a comparison between $\Delta \rho_{\rm rad}/\rho_{c,0}$ and $\Omega_{{\rm GW},0}$ (with subscript ,0 denoting the value today) from both before recombination (dashed) as well as from the whole cosmic history (solid).
The spectra of the $\Omega_{\rm GW,0}$ from before recombination can be understood as follows. The peak frequency is determined by the maximum GW frequency at the source and therefore is redshifted by $(1+z_\ast)$, where $z_\ast=1100$ is the redshift at recombination. The amplitude can be checked by performing the same kind of analysis as in Sec.~\ref{sec:SGWB_shape} by assuming the universe to be RD, which turns out to give a spectrum with very similar amplitude and power-law dependence $\fGW^{11/9}$ at high frequency (see the discussion in Appendix~\ref{app:OmegaGW_shape}). Thus the peak amplitude of $\Omega_{\rm GW,0}h^2\vert_{z_{\rm merge}>z_\ast}$ should be about $(1+z_\ast)^{11/9}$ smaller than the corresponding $\Omega_{\rm GW,0}$, which roughly matches the numerical results in the figure.
As a result, the SGWB from the dark binaries are less constrained by $\Delta N_{\rm eff}$. 

In addition to the GW emission, the DF mediator emission also contributes to $\DNeff$, which is shown in Fig.~\ref{fig:high_freq_signal} as dot-dashed curves. Note that the DF contribution to extra radiation energy dominates the GW contribution at lower frequencies, but they become more comparable at higher frequencies. This is because the binaries have smaller orbital semimajor axes at higher frequencies, so the GW emission becomes more important as illustrated in Eq.~(\ref{eq:dEdfs_complete}) and Fig.~\ref{fig:dEGWdfGWs}. Note that a calculation of the full constraint involves integrating and adding the whole GW and DF spectra. The relic density of the DF mediator also contributes to $\DNeff$, which can be neglected assuming the dark and visible sectors never thermalize and the dark sector is colder than the visible sector.

There are also existing or proposed direct detection experiments hunting for GWs in the MHz-GHz range. 
See~\cite{Aggarwal:2020olq,Ringwald:2020ist,Franciolini:2022htd} for reviews on possible detection approaches and their sensitivities; Ref.~\cite{Franciolini:2022htd} in particular discusses SGWBs at high frequencies from primordial BH binaries.
Unfortunately, even if the SGWB from MDM with a long-range DF can partially evade the constraint from $\Delta N_{\rm eff}$, the sensitivity of almost all of these proposed or existing methods are still not good enough to effectively constrain or detect the model parameter space, except for experiments based on the Gaussian beam (GB).
A GB consists of a conversion volume with an EM wave continuously propagating along its axis. A GW passing through the GB with a very similar frequency and parallel to the EM wave will convert to EM waves at first order in the GW strain amplitude $h_c$.
This significantly improves upon the GW sensitive compared to other experiments like single photon detection which have conversion probability proportional to $h^2_c$. Current GB experimental proposals usually operate at $\mc{O}(10){\rm GHz}$ and are expected to be sensitive to a GW characteristic strain of $h_c\sim 10^{-30}$~\cite{Li:2003tv,Ringwald:2020ist}. 
In Fig.~\ref{fig:high_freq_signal} we show the calculated SGWB before CMB and today for two benchmark values of $\rhodm^{1/4} = 120\uu\GeV$ and $\GpoG=10^4$, and compare with the setup of Ref.~\cite{Li:2003tv} where the GB is sensitive to a GW strain of $h_c=10^{-30}$ at $\fGW=10\uu{\rm GHz}$.~\footnote{These benchmark points have densities only a bit below that of black holes at these masses, and they are semirelativistic just before their merger. A full treatment would require a post-Newtonian simulation. This is beyond the scope of this work, and anyways other effects such as the GWs emitted after the merger have also been conservatively neglected due to the model dependence.}
Though by far the most promising direct detection method, this GB setup is more than an order of magnitude away from the benchmark points.
For the GB setup in Fig.~\ref{fig:high_freq_signal} to rule out the $\mdm=1.6\times 10^{-6}M_\odot$ benchmark point, it requires an improvement in $h_c$-sensitivity by (just) a factor of about three, while for the GB proposal in Ref.~\cite{Ringwald:2020ist} ($h_c=4\times 10^{-29}$ at $\fGW=40\uu{\rm GHz}$) it would be about $\mc{O}(100)$.
These upgrades could be reasonable in the near future through gyrotron improvements, see the end of Section 4 in~\cite{Ringwald:2020ist} for discussions on possible update directions.
Thus, GBs may start to rule out some model parameter space where the SGWB is compatible with the $\Delta N_{\rm eff}$ and not detectable by the lower-frequency experiments in the previous subsection.
On the other hand, more model building is required to understand models that populate this parameter space, since the benchmark points in the right panel of Fig.~\ref{fig:high_freq_signal} do not satisfy the model-dependent bounds in \eqref{eq:alpha_finite_density} and \eqref{eq:alpha-upper-vector}, based on the DQN models.

\section{Discussion and conclusions}
\label{sec:conclusion}

Throughout this work, the nearly massless mediator limit has been assumed. If the mediator mass is increased, it is possible that one or both of $\mmed \ll r^{-1}$ or $\mmed \ll \fGW$ may be violated for a portion of the inspiral phase of the binary. There are three possible cases to discuss.
In the case $\fGW \ll \mmed \ll r^{-1}$, the Kepler's law relation is modified, resulting in larger GW emission compared to the gravity-only case, but the DF emission is suppressed. In general, this suppression will lead to a substantially longer merger time than the massless mediator case because GW emission is much slower than DF emission in the massless mediator limit. This is true regardless if $f_\text{GW,min} \ll \mmed \ll f_\text{GW,max}$ or $f_\text{GW,max} \ll \mmed$---where $f_\text{GW,min,max}$ are the minimum and maximum frequency during the entire inspiral of the binary---because the lifetime is dominated by the lowest-frequency part of the inspiral. These two competing effects---the increased lifetime and suppressed DF emission coupled with enhanced GW emission---must be assessed on a case-by-case basis to determine if the overall SGWB spectrum increases or decreases. 
In the case $r^{-1} \ll \mmed \ll \fGW$, the binary is free to emit DF radiation, but the orbital frequency follows the ordinary Kepler's law determined by gravity. The DF emission leads to a shorter merger time than the gravity-only case, but the GW emission---which depends on the orbital frequency---is not enhanced at all until the orbital radius is sufficiently decreased to feel the attractive DF. Still, because the GW emission is always strongest at late times, if $r_\text{max}^{-1} \ll \mmed \ll r_\text{min}^{-1}$, an enhanced GW emission is possible along with an enhanced merger rate. On the other hand, compared to the massless mediator case, fewer binaries will form, and those that do form will form at later time with larger initial semimajor axis, because the attractive force in the early universe is reduced. 
Finally, if both $\mmed \gg r^{-1}$ and $\mmed \gg \fGW$ during a portion or all of the inspiral, the purely gravitational case is recovered. 
All of these cases require more detailed numerical treatment and are a topic for future work.

In addition to the massless mediator approximation, we have mostly assumed an attractive DF. Compared to the attractive DF assumption $\beta>1$ or even the gravity-only assumption $\beta=1$, a respulsive DF with $0<\beta<1$ would lead to a decreased GW emission per binary because the orbital frequency in Kepler's law is reduced and some of the orbital energy may still be emitted as DF radiation. Additionally, it will be harder for MDM objects to find each other and form binaries in the early universe because their attractive force is reduced.  This makes a repulsive force less promising for detectability.
Another possibility is a repulsive force with $\beta<0$ but a semi-massive mediator, where the mediator appears massless at longer ranges but massive at shorter ranges (though still longer than MDM radii). Then, binaries could form but would not merge. There would be a most stable distance between the two MDM objects, and as they emit radiation they would eventually come to rest at that position. While this would make for an interesting astrophysical system, clearly the GW emission is suppressed in this case.

In addition to their contribution to the SGWB, individual MDM inspirals or mergers could also be resolved. Indeed, the rate of binary mergers observed by HLVK already sets a constraint in the mass range around $\text{few} \times 10^{-1}$ to $10^3 M_\odot$ \cite{Hutsi:2020sol}. With a DF, the merger rate, signal strength, and waveform are also affected (the former two are discussed in this work; for the waveform, see, \eg, \cite{Diedrichs:2023foj}). As a result, the bound must be reinterpretted, which is beyond the scope of this work.

We have focused on dark binaries from MDM that forms well before matter-radiation equality.
For some MDM models, especially with some interactions in the dark sector similar to the electromagnetic interaction~\cite{Kouvaris:2015rea,Giudice:2016zpa,Maselli:2017vfi,Curtin:2019lhm,Curtin:2019ngc,Hippert:2021fch,Gross:2021qgx,Ryan:2022hku,Bai:2023mfi}, the MDM could be formed at a later time after the formation of large scale structure or galaxies. The properties, including the binary merger rates of dark stars or dark planets, are related to the details of the dark interactions. The additional long-range DF could diversify the detectable dark binary systems. For instance, there may be a sizable SGWB that is observable by other proposed high-frequency GW experiments besides the Gaussian beam, while being consistent with the $\Delta N_{\rm eff}$ constraints from the CMB observables.    

In summary, the presence of a long-range DF substantially modifies the formation, orbital dynamics, evolution, and radiation of MDM binaries due to its contribution to the binding force and addition of a new radiation mode. An attractive DF can increase or decrease the SGWB of merging binaries compared to the gravity-only case. Therefore, dark-charged MDM with planetary to solar masses can be detectable at much smaller abundances than are constrained by microlensing experiments. In particular, GWs at higher frequencies are the most enhanced compared to the gravity-only case, because an attractive force increases the orbital frequency and correspondingly the GW cutoff frequency from mergers. This makes them more detectable not only at aLIGO-aVIRGO or CE in the Hz-kHz range, but also at very high frequency detectors up to 100 GHz.

\subsubsection*{Acknowledgments}
The authors thank Jeff Dror and Harikrishnan Ramani for useful discussion. The work of YB is supported by the U.S. Department of Energy under the contract DE-SC-0017647. The work of SL is supported by the Area of Excellence (AoE) under the Grant No. AoE/P-404/18-3 issued by the Research Grants Council of Hong Kong S.A.R. The work of NO is partially supported by the National Science Centre, Poland, under research grant no.~2020/38/E/ST2/00243 and by the Arthur B. McDonald Canadian Astroparticle Physics Research Institute. The work of YB was performed in part at Aspen Center for Physics, which is supported by National Science Foundation grant PHY-2210452.

\appendix

\section{Dark quark nugget effective charge for the scalar-mediator model}
\label{app:scalar-model-effective-charge}

The finite-density effect of the DQNs could generate an in-medium potential for the scalar mediator. This effect could ``screen" the effective charge of a DQN to source the scalar field. In this Appendix, we keep a general form of the finite-density potential that contains linear, quadratic, and higher terms in $\phi$. To have an analytic calculation, we only keep the linear and quadratic terms. The effective potential is 
\beqa
V_{\rm eff}(\phi) = - \, a\,\phi\,\Theta(R - r) + \frac{1}{2}\,\left[ m_{\rm in}^2 \,\Theta(R-r) + \mmed^2 \,\Theta(r-R)\right]\,\phi^2 ~. 
\eeqa
Here, $\Theta(x)$ is the Heaviside step function; $a$ is proportional to the fermion number density and the Yukawa coupling; $m_{\rm in}$ is the scalar mass inside the DQN; and $\mmed$ is the scalar mass outside the DQN (without the medium effect). For a spherical configuration, the classical equation of the scalar mediator is
\beqa
\partial^2_r \phi + \frac{2}{r} \partial_r \phi = \frac{\partial V_{\rm eff}(\phi)}{\partial \phi} ~,
\eeqa
with the effective potential $V_{\rm eff}(\phi)$ as a summation of $V_0$ in \eqref{eq:yukawa} and $V_1$ in \eqref{eq:potential-V1}. Outside the DQN with $\mmed^2 >0$ and the boundary condition $\phi(\infty) = 0$, the solution is $\phi_{\rm out} = c_1 e^{-\mmed \, r}$. Inside the DQN with $m_{\rm in}^2 >0$ and with the Neumann boundary condition $\phi'(r=0)=0$, the solution is $\phi_{\rm in} = a/m_{\rm in}^2 + c_2 (e^{- m_{\rm in}\,r} - e^{m_{\rm in}\,r})/r$. After using the continuity boundary conditions $\phi_{\rm in}(R) = \phi_{\rm out}(R)$ and $\phi'_{\rm in}(R) = \phi'_{\rm out}(R)$, one has 
\beqa
\frac{q_{\rm eff}\,y}{4\pi} =c_1 &=& a\, \frac{e^{\mmed\,R}[m_{\rm in}\,R\,\cosh{(m_{\rm in}\,R)} - \sinh{(m_{\rm in}\,R)}]}{m_{\rm in}^2[\mmed\,\sinh{(m_{\rm in}\,R)} + m_{\rm in}\,\cosh{(m_{\rm in}\,R)}]} \,, \\
c_2 &=& a\, \frac{\mmed\,R + 1}{2\,m_{\rm in}^2[m_{\rm in}\,\cosh{(m_{\rm in}\,R)} + \mmed\,\sinh{(m_{\rm in}\,R)}]} ~. 
\eeqa
In the limit of $m_{\rm in} R, \mmed R \ll 1$, one has $c_1 = a\,R^3/3$ or $q_{\rm eff}\,y = a\,4\pi R^3/3$ as anticipated for a spherically symmetric constant source. In the limit with only $\mmed R \ll 1$, one has 
\beqa
c_1 = a\,\frac{\left[m_{\rm in}\,R - \tanh(m_{\rm in}\,R)\right]}{m_{\rm in}^3} \xrightarrow{m_{\rm in} R\rightarrow\infty}a\,\frac{R^3}{(m_{\rm in}\,R)^2} ~,
\eeqa
which is suppressed by $1/(m_{\rm in}\,R)^2$ when the medium-induced scalar mass is much heavier than $1/R$. So, to keep a large effective charge to source the scalar mediator, one may required $m_{\rm in}\,R < 1$. 

For the case with a negative effective mass inside or $m_{\rm in}^2 < 0$, one can define $\overline{m}_{\rm in}^2 = - m_{\rm in}^2$. The outside solution stays the same, while the inside solution is modified to be $\phi_{\rm in} =-a/\overline{m}_{\rm in}^2 + c_2 (e^{- i\,\overline{m}_{\rm in}\,r} - e^{i\,\overline{m}_{\rm in}\,r})/r$. After matching the boundary conditions and taking the limit of $\mmed R \ll 1$, the formula for $c_1$ is similar to the previous case with a positive $m_{\rm in}^2$ and is given by
\beqa
c_1 = -\,a\,\frac{\left[\overline{m}_{\rm in}\,R - \tanh(\overline{m}_{\rm in}\,R)\right]}{\overline{m}_{\rm in}^3} \xrightarrow{\overline{m}_{\rm in} R\rightarrow\infty} \,-\,a\,\frac{R^3}{(\overline{m}_{\rm in}\,R)^2} ~.
\eeqa

\section{Radiation emission from a binary with a massless dark force mediator}
\label{app:emission_rate}

\subsection{GW emission}

Here, we review the ordinary GW emission from a binary with no additional long-range forces. For further review, see, \eg, \cite{Maggiore:08textbook}. The quadruple radiation depends on the quadruple
\begin{equation}
    Q^{ij} \equiv M^{ij} - \frac{1}{3} \delta^{ij} M_{kk} \, ,
\end{equation}
where the second mass moment
\begin{equation}
    M^{ij} = \int d^3x \, T^{00} x^i x^j \, ,
\end{equation}
with $T^{00}$ the energy density from the stress-energy tensor. To lowest order in velocity $v$, $T^{00} \approx \rho$ the mass density. The total radiated quadruple power is then
\begin{equation}
    \dot{E}_\text{GW} = \frac{G}{5} \langle \dddot{Q}_{ij} \dddot{Q}_{ij} \rangle = \frac{G}{5} \langle \dddot{M}_{ij} \dddot{M}_{ij} - \frac{1}{3} (\dddot{M}_{kk})^2 \rangle \, ,
\end{equation}
where dots reflect derivatives with respect to time $t$.
For a binary, the radiated power averaged over its orbital period is
\begin{equation}
\label{eq:PGW}
    P_\text{GW} = \langle \dot{E}_\text{GW} \rangle = \frac{32 G G'^3 \eta^2 m^5}{5 a^5 (1-e^2)^{7/2}} \left(1 + \frac{73}{24}e^2 + \frac{37}{96}e^4 \right) \, .
\end{equation}
Note that the single power of $G$ comes from the gravitational coupling for the GW emission, while the other three powers of $G'$ come from the orbital trajectory in the mass density in $\dddot{M}_{ij}$.

Meanwhile, the angular moment radiated in quadruple GWs is
\begin{equation}
    \dot{L}_\text{GW}^i = \frac{2 G}{5} \epsilon^{ikl} \langle \ddot{Q}_{ka} \dddot{Q}_{la} \rangle \, .
\end{equation}
For a binary, this is
\begin{equation}
    \dot{L}_\text{GW} = \frac{32 G G'^{5/2} \eta^2 m^{9/2}}{5 a^{7/2} (1-e^2)^2} \left(1+\frac{7}{8}e^2\right) \, .
\end{equation}

Consider the time for a binary with initial orbital parameters $a_0$ and $e_0$ to merge if GW emission is the dominant/only source of radiation.

To estimate the merger time, first obtain $\dot{a}$ and $\dot{e}$ from $\dot{E}_\text{GW}$ and $\dot{L}_\text{GW}$ as [using (\ref{eq:energy}) and $e^2 = 1+ \frac{2 E L^2}{G'^2 m^5 \eta^3}$]
\begin{align}
    \dot{a} &= - \frac{2 a^2}{G' m^2 \eta} \dot{E}_\text{GW} = - \frac{64 G G'^2 \eta m^3}{5 a^3 (1-e^2)^{7/2}} \left(1 + \frac{73}{24}e^2 + \frac{37}{96}e^4 \right) \, ,
    \\
    \dot{e} &= - \frac{\dot{E}_\text{GW} L_\text{GW}^2 + 2 E_\text{GW} L_\text{GW} \dot{L}_\text{GW}}{G'^2 m^5 \eta^3 e} = -\frac{304 G G'^2 \eta m^3 e}{15 a^4 (1-e^2)^{5/2}} \left(1+\frac{121}{304}e^2\right) \,.
\end{align}
These can be combined to get $da/de$, which can be integrated analytically to obtain
\begin{equation}
\begin{aligned}
    \label{eq:aeGW}
    a(e) &= a_0 \frac{g(e)}{g(e_0)} \, ,
    \\
    g(e) &= \frac{e^{12/19}}{1-e^2} \left(1+\frac{121}{304} e^2\right)^{870/2299} \, .
\end{aligned}
\end{equation}
Noting that $e \to 0$ as $a \to 0$, the merger time is obtained from
\begin{align}
    \tau &= \int_{e_0}^0 de \left(\frac{de}{dt}\right)^{-1} = \frac{5 a_0^4 (1-e_0^2)^{7/2}}{256 G G'^2 \eta m^3} G(e_0) \, ,
    \\
    G(e_0) &= \frac{48}{19 g^4(e_0) (1-e_0^2)^{7/2}} \int_0^{e_0} de \frac{g^4(e)(1-e^2)^{5/2}}{e(1+\frac{121}{304}e^2)} \, .
\end{align}
The function $G(e_0)$ is monotonic in $e_0 \in [0,1)$ and satisfies $G(0)=1$ and $G(1)=768/425$ (note that for the $e \to 1$ limit, $5/256 \times 768/425 = 3/85$).

\subsection{Dark force mediator emission}

\subsubsection{Vector mediator}
The dipole radiation of the massless dark mediator is related to the dipole moment $\bm{p} = g q_1 \bm{x}_1 + g q_2 \bm{x}_2$
\begin{equation}
    \dot{E}_\text{EM} = -\frac{2}{3} \frac{1}{4 \pi } \ddot{p}^2 = -\frac{2}{3} G'^2 (g q_1 m_2 - g q_2 m_1)^2 \frac{1}{r^4} \, .
\end{equation}
After time-averaging $1/r^4$ over a period of the orbit, the average energy loss is
\begin{equation}\label{eq:E_dot}
\begin{aligned}
    P_\text{EM} = \langle \dot{E}_\text{EM} \rangle &= \frac{G'^2}{12 \pi} (g q_1 m_2 - g q_2 m_1)^2 \frac{1}{a^4} \frac{2+e^2}{(1-e^2)^{5/2}}
    \\
    &= \frac{G G'^2}{12 \pi} \eta^2 m^4 \left(\frac{g q_1}{\sqrt{G}m_1}- \frac{g q_2}{\sqrt{G}m_2}\right)^2 \frac{1}{a^4} \frac{2+e^2}{(1-e^2)^{5/2}} \, .
\end{aligned}
\end{equation}
For the same-mass opposite-charge case $q_1 = -q_2$ and $m_1 = m_2$, the middle term in the last line (the squared difference in the charge-to-mass ratios) is $16 \pi |\alpha|$.

The loss of angular momentum goes as \cite{Liu:2020cds},
\begin{equation}
    \langle \dot{L}_\text{EM} \rangle = \frac{G'^{3/2}(g q_1 m_2 - g q_2 m_1)^2}{6 \pi a^{5/2} (1-e^2) \sqrt{m}} \, .
\end{equation}
As in the GW case [using (\ref{eq:energy}) and $e^2 = 1+ \frac{2 E L^2}{G'^2 m^5 \eta^3}$]
\begin{align}
    \dot{a} &= - \frac{2 a^2}{G' m^2 \eta} \dot{E}_\text{EM} = - \frac{G G'}{6 \pi} \eta m^2 \left(\frac{g q_1}{\sqrt{G}m_1}- \frac{g q_2}{\sqrt{G}m_2}\right)^2 \frac{1}{a^2} \frac{2+e^2}{(1-e^2)^{5/2}}  \, ,
    \\
    \dot{e} &= - \frac{\dot{E}_\text{EM} L_\text{EM}^2 + 2 E_\text{EM} L_\text{EM} \dot{L}_\text{EM}}{G'^2 m^5 \eta^3 e} = - \frac{G G' m^2 \eta e}{4 \pi a^3 (1-e^2)^{3/2}} \left(\frac{g q_1}{\sqrt{G}m_1}- \frac{g q_2}{\sqrt{G}m_2}\right)^2 \, .
    \label{eq:dote_EM}
\end{align}
These can be combined to obtain $da/de$, which can be integrated analytically to obtain
\begin{equation}
\begin{aligned}
    \label{eq:aeEM}
    a(e) &= a_0 \frac{g(e)}{g(e_0)} \, ,
    \\
    g(e) &= \frac{e^{4/3}}{1-e^2} \, .
\end{aligned}
\end{equation}
Noting that $e \to 0$ as $a \to 0$, the merger time is obtained from
\begin{align}
    \tau &= \int_{e_0}^0 de \left(\frac{de}{dt}\right)^{-1} 
    \\
    & = \frac{4 \pi a_0^3}{G G' \eta m^2} \left(\frac{g q_1}{\sqrt{G}m_1}- \frac{g q_2}{\sqrt{G}m_2}\right)^{-2} \frac{1}{g^3(e_0)} \int_0^{e_0} de \frac{g^3(e)(1-e^2)^{3/2}}{e}
    \\\label{eq:tau_EM}
    &= \frac{4 \pi a_0^3}{G G' \eta m^2} \left(\frac{g q_1}{\sqrt{G}m_1}- \frac{g q_2}{\sqrt{G}m_2}\right)^{-2} \frac{(1-e_0^2)^{5/2} (1-\sqrt{1-e_0^2})^2}{e_0^4} \, .
\end{align}
The last fraction containing the $e_0$ goes to 1/4 as $e_0 \to 0$.

\subsubsection{Scalar mediator}

The scalar mediator emission is nearly identical to the vector mediator emission, but both $\langle \dot{E} \rangle$ and $\langle \dot{L} \rangle$ are smaller by a factor of two in the massless mediator limit (see \cite{Krause:1994ar} for $\langle \dot{E} \rangle$ and below for $\langle \dot{L} \rangle$). As a result, $\dot{e}$ is smaller by a factor of two compared to Eq.~(\ref{eq:dote_EM}), and the lifetime for given initial $a_0$ and $e_0$ is larger by a factor of two compared to Eq.~(\ref{eq:tau_EM}). The change in lifetime can be absorbed by $\bar{\tau}$ in (\ref{eq:taubar}), which would become larger by a factor of two. The expression for $dE_\text{GW}/df_\text{GW}$ is also increased by a factor of two relative to the EM case when the DF emission dominates, though it is unchanged when GWs dominate. In $\Omega_\text{GW}$ in (\ref{eq:GW_spec2}) as well as precursor equations like (\ref{eq:merger-rate-differential}), the $n_\MDM/4$ becomes instead $n_\MDM/2$ for the case where all MDM carries the same (attractive) scalar charge. Thus, the overall $\Omega_\text{GW}$ integration gives a very similar result for the scalar mediator case as the vector mediator case.

To show the equivalence between the vector and scalar emission we need to show that their emission power $\dot{E}$ and angular momentum loss $\dot{L}$ are the same up to a redefinition of the coupling.
It has been shown that with a massless mediator the emission power from the two types of mediators differs by a factor of two~\cite{Krause:1994ar,Alexander:2018qzg}, and therefore we only need to show that the corresponding $\dot{L}$ differ by the same factor.
The angular momentum loss from the vector mediator has been calculated in Ref.~\cite{Liu:2020cds} in terms of the time derivatives of the dipole, while no result has been given for the scalar mediator, partially due to the not-so-straightforward interpolation.
For our purpose, however, we may take a detour and use the formalism in Ref.~\cite{Krause:1994ar} to show that the angular momentum loss in the vector mediator model is twice as large as the scalar mediator model.

The angular momentum loss is calculated as~\cite{Liu:2020cds}, 
\begin{align}
\dot{L}=\dfrac{dJ}{dt}=-\int r^2 d\Omega\uu  j({\bf r})\,.
\end{align}
We will keep only the leading order terms in the multipole expansion of the source terms, as well as only the leading order $1/r$ terms.
For the (real) scalar mediator model, the angular momentum density $j$ is
\begin{align}
j^i=-\dot{\Phi}\epsilon^{ijk}r^j\partial^k \Phi\,.
\end{align}
Using the approach of~\cite{Krause:1994ar}, we expand the scalar field into Fourier modes as
\begin{align}
\Phi_{\rm rad}=\sum_{\lvert n\rvert>n_0}I_n(\hat{r})\dfrac{\exp\left[i(k_n\, r-n\,\omega_0\, t)\right]}{r}\equiv\sum_{\lvert n\rvert>n_0}I_n(\hat{r})\dfrac{E_n}{r}\,,
\end{align}
where
\begin{align}
I_n=f_S\int d^3x^\prime \exp\left(-ik_n\hat{r}\cdot\bv{x}^\prime\right)\rho_n(\bv{x}^\prime)\,,\quad \rho=\sum_n\rho_ne^{-i\uu n\uu\omega_0\uu t}\,,
\end{align}
with $f_S$ as the effective coupling and $\hat{r}$ the unit vector along the direction of $\bm{r}$, whose derivative and angular integration satisfy
\begin{gather}
\partial^a \hat{r}^b=\dfrac{1}{r}(\delta^{ab}-\hat{r}^a\hat{r}^b)\,,\\
\int d\Omega \hat{r}^a \hat{r}^b=\dfrac{4\pi}{3}\delta^{ab}\,,\quad \int d\Omega \hat{r}^a \hat{r}^b \hat{r}^c \hat{r}^d=\dfrac{4\pi}{15}(\delta^{ab}\delta^{cd}+\delta^{ac}\delta^{bd}+\delta^{ad}\delta^{bc})\,.
\end{gather}
Thus,
\begin{align}
\dot{L}=&+\int d\Omega \sum_{\lvert n\rvert,\lvert m\rvert}(-in\omega_0)I_n \epsilon^{ijk}r^j \left((-i\uu k_m)f_S(\partial^k \hat{r})\cdot\int d^3 y^\prime \bv{y^\prime} e^{-ik_m\hat{r}\cdot\bv{y^\prime}}
\rho_m(\bv{y}^\prime)\right)\langle E_nE_m\rangle \nonumber\\
\approx &\int d\Omega \sum_{\lvert n\rvert}(-in\omega_0)\left(f_S\int d^3 x^\prime (-ik_n\hat{r}^b (x^\prime)^b)\rho_n(\bv{x}^\prime)\right) \epsilon^{ijk}\hat{r}^j \left((i\uu k_n)f_S\int d^3 y^\prime (y^\prime)^k \rho_{-n}(\bv{y}^\prime)\right) \nonumber\\
=&\sum_{\vert n\rvert > n_0}\dfrac{4\pi}{3}\left(-i\uu n^3\omega^3_0\left(1-\dfrac{n^2_0}{n^2}\right)f^2_S\epsilon^{ijk}p^j_n p^k_{-n}\right)\,,\label{eq:Ldot_scalar_portal}
\end{align}
where
\begin{align}
\bv{p}_n=\int d^3 x^\prime \bv{x}^\prime \rho_n(\bv{x}^\prime)\,.
\end{align}
Note that for the exponential inside $I_n$ in the first line of (\ref{eq:Ldot_scalar_portal}), we have expanded to leading order, as the total charge is conserved and therefore $\int d^3x^\prime \rho_n$ with $n \neq 0$ vanishes.

For the vector mediator model, the angular momentum density is~\footnote{Note that the expression in the Appendix of~\cite{Liu:2020cds} uses temporal gauge while the results in~\cite{Krause:1994ar} uses Lorentz gauge.}
\begin{align}\label{eq:Ji_Lorentz_gauge}
j^i=\epsilon^{ijk}\left(-\partial_0\uu A_l \uu r_j \partial_k A^l+\partial_l\uu A_0 \uu r_j \partial_k A^l+A_j\partial_0A_k-A_j\partial_k A_0\right)\,.
\end{align}
We use the metric $g_{\mu\nu}={\rm diag}\{-1,1,1,1\}$ further on. Again,~\cite{Krause:1994ar} suggests
\begin{align}
A^\alpha=\sum_{\vert n\vert}I^\alpha_n(\hat{r})\dfrac{E_n}{r}\,,\quad I^\alpha_n(\hat{r})=f_V \int d^3x^\prime e^{-ik_n\hat{r}\cdot\bv{x}^\prime}J^\alpha_n(\bv{x}^\prime)\,,
\end{align}
where $f_V$ is the gauge coupling and $J^\alpha$ is the current.
To the zeroth order, 
\begin{align}
J^i_n(\bv{x}^\prime)\approx -i\uu n\omega_0 (x^\prime)^i\rho_n(\bv{x}^\prime)\,,\quad
I^i_n\approx -i\uu n\omega_0 f_V\int d^3 x^\prime (x^\prime)^i \rho_n(\bv{x}^\prime)=-i\uu n\omega_0 f_V p_n^i\,.\label{eq:0th_order_vector_In}
\end{align}
For the zeroth component, Lorentz gauge implies that $I^0_n=\sqrt{1-(n_0/n)^2}\uu\hat{r}\cdot\bv{I}_n(\bv{x}^\prime)$ to leading order in $1/r$, and therefore
\begin{align}
\partial_k A_0=&-\partial_k\left(\sum_{\lvert n\rvert}\sqrt{1-\left(\dfrac{n_0}{n}\right)^2}\hat{r}^a I^a_n \dfrac{E_n}{r}\right)\,,
\end{align}
where the negative sign in the first line comes from the metric.

The leading-order contributions to the angular momentum loss in the vector mediator model come from the last two terms in Eq.~\eqref{eq:Ji_Lorentz_gauge}.
A straightforward calculation shows that for the third term, 
\begin{align}
\dfrac{dJ^i_{(3)}}{dt}=-\int r^2d\Omega\uu \epsilon^{ijk}A_j\partial_0 A_k=\sum_{\lvert n\rvert}4\pi\left(-i\uu n^3\omega^3_0f^2_V\epsilon^{ijk}p^j_n p^k_{-n}\right)\,.
\end{align}
For the fourth term,
\begin{align}
\dfrac{dJ^i_{(4)}}{dt}=-\int r^2d\Omega\uu\epsilon^{ijk}(-A_j\partial_k A_0)=\sum_{\lvert n\rvert}\dfrac{4\pi}{3}\left(i \, n^3\left(1-\dfrac{n^2_0}{n}\right)\omega^3_0 \,f^2_V\,\epsilon^{ijk}\,p^j_n\, p^k_{-n}\right)\,,
\end{align}
where in the derivation we have discarded higher-order terms in $1/r$.
Note that the contribution from these two terms are at the zeroth order of the multiple expansion, \ie, taking $e^{-ik_n\hat{r}\cdot\bv{x}^\prime}=1$ as in Eq.~\eqref{eq:0th_order_vector_In}.
At this order, for the first term of Eq.~\eqref{eq:Ji_Lorentz_gauge}, 
\begin{align*}
\dfrac{dJ^i_{(1)}}{dt}=&\int r^2 d\Omega\uu\epsilon^{ijk} \partial_0 A_l \uu r_j \partial_k A^l\\
=&\int d\Omega\sum_{\lvert n\rvert,\lvert m\rvert}\epsilon^{ijk}\left[(-i\uu n\,\omega_0)I^l_n\right]r^j\left[f_V\dfrac{-i\uu k_m}{r}(\delta^{ka}-\hat{r}^k\hat{r}^a)\int d^3 y^\prime (y^\prime)^a e^{-ik_m\hat{r}\cdot\bv{y}^\prime}J^l_m(\bv{y}^\prime)\right.\\
&\ \ \ \left.+I^l_m\left(i\uu k_m-\dfrac{1}{r}\right)\hat{r}^k\right]\langle E_nE_m\rangle\,.
\end{align*}
This entire expression vanishes: the first term in the square brackets contains an odd number of $\hat{r}$ unit vectors and vanishes upon angular integration, while the second term is zero because of the Levi-Civita tensor $\epsilon^{ijk}$.
The second term of~\eqref{eq:Ji_Lorentz_gauge} similarly vanishes.
Summing up the (nonzero) third and fourth terms and comparing the summation with Eq.~\eqref{eq:Ldot_scalar_portal} noticing that for a massless mediator $n_0=0$~\cite{Alexander:2018qzg}),
we see that the angular momentum loss of the vector mediator model is twice of that of the scalar mediator model when $f_S=f_V$.

\section{More details on the shape of the GW spectra}\label{app:OmegaGW_shape}

The GW spectra that we calculate clearly show a change of power-law index across their frequency bands, which we analyze below by checking the behavior of the $\Omega_\text{GW}$ integrand in (\ref{eq:GW_spec2}).
For concreteness we use the blue curve in Fig.~\ref{fig:spectra_shape} as a benchmark example, which is calculated with $\mdm=10^{-7} M_\odot$, $f=0.01$, $G^\prime/G=10^4$ and $\rho=1\,{\rm GeV}^4$.

We start from the high frequency regime close to the cut-off of the spectrum, \ie, the $\fGW\gtrsim 10^{-3}$ Hz region in our benchmark case.
As illustrated in the main text, in this regime the EM emission dominates the GW emission and the orbital eccentricity has been softened from $e\sim 1$.
Taking the approximation $t=\tau$ and $e=0$, assuming matter domination, and ignoring the GW emission term in the denominator of Eq.~\eqref{eq:dEdfs_complete} one has
\begin{align}
\dfrac{dE}{d\fGWs}\approx \left(\dfrac{2}{3H_0\sqrt{\Omega_{\rm M}}\tau}\right)^{2/9} \dfrac{4\times 2^{1/3}\pi^{4/3}G^{\prime\uu 4/3}\mdm^{7/3} \fGW^{1/3}}{5(\GpoG-1)}\,.
\end{align}
Now together with the rest of the terms, the integrand of Eq.~\eqref{eq:GW_spec2} has a $\tau$-dependence of $A\uu\tau^{-13/18} e^{-B\uu\tau^{1/4}}$, and the inner $\tau$-integration evaluates to
\begin{align}
\label{eq:inner_tau_int}
&\int^{\tau_{\rm max}}_{\tau_{\rm min}} d\tau\uu A\uu \tau^{-13/18} e^{-B\uu\tau^{1/4}}\approx 4A\uu B^{-10/9}\uu \Gamma\left(\frac{10}{9},B\uu\tau^{1/4}_{\rm min}\right)\,,\\
A&=\dfrac{2^{\frac{1}{18}}3^{\frac{5}{18}}\pi^{\frac{5}{6}}c_2(1+z_{\rm eq})^{3/2} f^2\GpoG^{10/3} G^{11/6} H^{7/9}_0 \mdm^{17/6}\Omega^{1/2}_{\rm DM}}{5c^{3/2}_1(\GpoG-1)^{1/2}\Omega^{1/9}_{\rm M}}\dfrac{e^3_0}{(1-e^2_0)^{11/4}\left(1-\sqrt{1-e^2_0}\right)}\fGW^{1/3}\,,\\
B&=\dfrac{3^{1/4}c_2 (1+z_{\rm eq})^{3/4} f\uu \sqrt{H_0}\uu \GpoG \uu (G^\prime-G)^{1/4}\mdm^{1/4}\Omega^{1/4}_{\rm DM}}{(2\pi)^{1/4}c^{3/4}_1}\dfrac{e_0}{(1-e^2_0)^{9/8}\left(1-\sqrt{1-e^2_0}\right)^{1/2}}\,.
\end{align}
We further approximate $\Gamma\left(\frac{10}{9},z\right)\approx z^{1/9}\exp(-z)$ where $z=B\uu \tau^{1/4}_{\rm min}$. With this, Eq.~(\ref{eq:inner_tau_int}) is proportional to $(1-e^2_0)^{-13/8}\exp[-C/(1-e^2_0)^{9/8}]$ (where $C$ is positive and independent of $e_0$) and therefore is peaked at an $e_0$ very close to one. We can then evaluate the outer integration with respect to $e_0$ using the saddle-point approximation (after setting powers of $e_0$ and $1-\sqrt{1-e^2_0}$ to 1 and keeping only powers of $1-e^2_0$). This leads to the expression in Eq.~\eqref{eq:OmegaGW_high}
\begin{align}
\Omega_{\rm GW,UV}\approx \dfrac{4\times 2^{\frac{5}{6}}\times 13^{\frac{1}{18}} \pi^{\frac{59}{36}} f^{\frac{13}{9}} G^{\frac{55}{36}}\GpoG^{\frac{67}{36}}(1+z_{\rm eq})^{\frac{1}{3}}\rhodm^{\frac{1}{12}} H^{\frac{1}{9}}_0 \mdm^{\frac{13}{9}} \Omega^{\frac{10}{9}}_{\rm DM}}{45\times 3^{7/36} e^{13/9} c^{1/3}_1 c^{5/9}_2 (\GpoG-1)^{8/9} \Omega^{1/18}_{\rm M}}\fGW^{7/6}\,,\nonumber
\end{align}
which we refer to as the UV piece of the spectrum. The density $\rhodm$ is involved through $\tau_{\rm min}$, which in this case is determined by the maximum frequency at the source, \ie, the RHS of the constraint~\eqref{eq:tau_constraint_3}. Note that the power-law index of $7/6$ depends on our underlying MD assumption. 
Repeating the exercise above with an RD assumption gives
\begin{align}
\Omega_{\rm GW}\approx \dfrac{4\times 2^{\frac{11}{18}}\times 13^{\frac{1}{18}} \pi^{\frac{5}{3}} f^{\frac{13}{9}} G^{\frac{3}{2}}\GpoG^{\frac{11}{6}}(1+z_{\rm eq})^{\frac{1}{3}}\rhodm^{\frac{1}{18}} H^{\frac{1}{9}}_0 \mdm^{\frac{13}{9}} \Omega^{\frac{10}{9}}_{\rm DM}}{45\times 3^{1/18} e^{13/9} c^{1/3}_1 c^{5/9}_2 (\GpoG-1)^{8/9} \Omega^{1/18}_{\rm R}}\fGW^{11/9}\,.\nonumber
\end{align}

At smaller $\fGW$, the mergers largely happen when the universe is radiation dominated, and the orbital eccentricity is largely not softened from one. Meanwhile, the binaries merge swiftly after their decoupling, which means that  $t_{\rm dec}\gg \tau$ and
\begin{align}
\label{eq:taumin2}
\tau_{\rm min}\approx\left(\dfrac{3\pi \fGW^2}{G^\prime \rhodm}\right)^2\dfrac{2\pi c^3_1(1+z_{\rm eq})h(e_0)}{3H^2_0 G \mdm(\GpoG-1)\Omega_{\rm DM}}\,.
\end{align}
With these changes, to a good approximation we will take $e=1$ in Eq.~\eqref{eq:dEdfs_complete} (while note that we make the replacement that $a(1-e^2)=a_0 (e/e_0)^{4/3}(1-e^2_0)$ from (\ref{eq:aeEM}) such that the EM process contribution in the denominator doesn't immediately vanish).
Then the competition between the EM emission and the GW emission in the denominator of~\eqref{eq:dEdfs_complete} determines the spectral shape.
When the EM emission overpowers the GW emission, we may ignore the GW terms in the denominator. Then, with an RD assumption the integrand of~\eqref{eq:GW_spec2} behaves like $A\uu\tau^{-3/4} e^{-B\uu\tau^{1/4}}$, and the $\tau$-integration evaluates to 
\begin{align}\label{eq:e0-integrand-EM}
&\int^{\tau_{\rm max}}_{\tau_{\rm min}} d\tau\uu A\uu\tau^{-3/4} e^{-B\cdot \tau^{1/4}}=4A\uu B^{-1}\exp(-B\uu \tau^{1/4}_{\rm min})\,,\\
A&=\dfrac{85\pi^{\frac{1}{12}}c_2 (1+z_{\rm eq})^{17/12} f^2\GpoG^{10/3} G^{19/12} H^{7/6}_0 \mdm^{31/12}\Omega^{7/12}_{\rm DM}}{48\times 2^{\frac{7}{12}}\times 3^{\frac{5}{12}}c^{7/4}_1(\GpoG-1)^{3/4}}\dfrac{e^{10/3}_0}{(1-e^2_0)^{25/8}\left(1-\sqrt{1-e^2_0}\right)^{1/2}}\fGW^{-1/3}\,,\\
B&=\dfrac{3^{1/4}c_2 (1+z_{\rm eq})^{3/4} f\uu \sqrt{H_0}\uu \GpoG \uu (G^\prime-G)^{1/4}\mdm^{1/4}\Omega^{1/4}_{\rm DM}}{(2\pi)^{1/4}c^{3/4}_1}\dfrac{e_0}{(1-e^2_0)^{9/8}\left(1-\sqrt{1-e^2_0}\right)^{1/2}}\,.
\end{align}
In terms of $(1-e_0^2)$, Eq.~(\ref{eq:e0-integrand-EM}) behaves as $(1-e^2_0)^{-2}\exp[-C/(1-e^2_0)^{1/2}]$. Performing the outer $e_0$-integration with a saddle-point approximation, the GW spectra in this situation is
\begin{align}
\label{eq:OmegaGW_middle}
\Omega_{\rm GW,middle}\approx \dfrac{85\uu G^{7/3}\GpoG^{4/3}H^{2/3}_0 \mdm^{4/3}\Omega^{4/3}_{\rm DM}\rhodm}{3^{5/3}\times 2^{5/6} \pi^{1/6}(1+z_{\rm eq})^{4/3}c_1 c^2_2 e^4(\GpoG-1) }\fGW^{-4/3}\,.
\end{align}
Note that the power-law index on $\fGW$ is now negative, and this corresponds to the region $10^{-3}{\rm Hz}\lesssim \fGW\lesssim 10^{-2}{\rm Hz}$ of our benchmark point, which we refer to as the middle piece of the spectrum.

\begin{figure}
    \centering
    \includegraphics[width=0.6\textwidth]{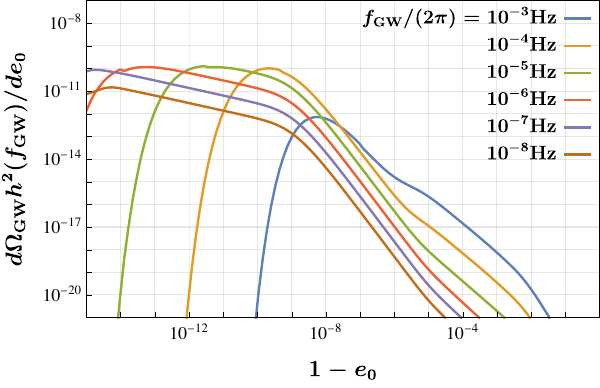}
    \caption{The integrand of the outer $e_0$-integration as a function of $1-e_0$ for various $\fGW$. The horizontal axis is $1-e_0$, \and hence $e_0\to 1$ to the left. From right to left the curves are for $\fGW/(2\pi)=10^{-3}, 10^{-4}, ..., 10^{-8}{\rm Hz}$, and the model parameters are the same as the blue curve in Fig.~\ref{fig:spectra_shape}.
    For $\fGW/(2 \pi)\lesssim 10^{-5}{\rm Hz}$, the integrand starts to have a knee at $1-e_0\sim 10^{-9}$, which is due to the competition between the EM and GW emission. At smaller $1-e_0$ the EM emission dominates.}
    \label{fig:integrand_e0}
\end{figure}

At even smaller $\fGW$ the competition between the EM emission and GW emission becomes more complicated. Even though the GW emission is the main contributor to $\Omega_{\rm GW}$, it overpowers the EM emission only in the region where $e_0$ is very close to one. Fig.~\ref{fig:integrand_e0} shows the integrands of the $e_0$-integration for various $\fGW$ as a function of $1-e_0$, where there is a knee for most curves around $(1-e_0)\sim 10^{-9}$, to whose left (where $e_0$ is closer to 1) the GW emission dominates.
In this region the EM emission term in the denominator of~\eqref{eq:dEdfs_complete} may be ignored, and the $\tau$-integration evaluates to
\begin{align}\label{eq:e0-integrand-GW}
&\int^{\tau_{\rm max}}_{\tau_{\rm min}} d\tau\uu A\uu\tau^{-5/12} e^{-B\uu \tau^{1/4}}=4A\uu B^{-7/3}\Gamma\left(\dfrac{7}{3},B\uu\tau^{1/4}_{\rm min}\right)\,,\\
A&=\dfrac{\pi^{\frac{1}{12}}c_2 (1+z_{\rm eq})^{17/12} f^2\GpoG^{8/3}(\GpoG-1)^{7/12} G^{5/4} H^{7/6}_0 \mdm^{9/4}\Omega^{7/12}_{\rm DM}}{4\times 2^{\frac{11}{12}}\times 3^{\frac{5}{12}}c^{7/4}_1}\dfrac{e^{10/3}_0}{(1-e^2_0)^{7/6}\left(1-\sqrt{1-e^2_0}\right)^{7/6}}\fGW^{-1/3}\,,\\
B&=\dfrac{3^{1/4}c_2 (1+z_{\rm eq})^{3/4} f\uu \sqrt{H_0}\uu \GpoG \uu (G^\prime-G)^{1/4}\mdm^{1/4}\Omega^{1/4}_{\rm DM}}{(2\pi)^{1/4}c^{3/4}_1}\dfrac{e_0}{(1-e^2_0)^{9/8}\left(1-\sqrt{1-e^2_0}\right)^{1/2}}\,.
\end{align}
Equating the expressions in~\eqref{eq:e0-integrand-EM} and~\eqref{eq:e0-integrand-GW} determines the position of the knee, which we denoted $e_{0,{\rm th}}$ in the main text.
Since we are working in the IR regime where $\fGW$ is small, $B\uu\tau^{1/4}_{\rm min}\ll 1$, in which case
\begin{align}\label{eq:e0th}
1-e^2_{0,{\rm th}}=\left(\dfrac{85c^{4/3}_2(1+z_{\rm eq})f^{4/3}\GpoG^2(G\uu H_0\mdm)^{2/3}\Omega^{1/3}_{\rm DM}}{4\times 3^{2/3}\pi^{1/3}\Gamma\left(\frac{7}{3}\right)c_1(\GpoG-1)}\right)^{3/5}\,.
\end{align}
After making the approximation that $\Gamma(\frac{7}{3},B\uu\tau^{1/4}_{\rm min})\approx\Gamma(\frac{7}{3})$ in~\eqref{eq:e0-integrand-GW}, the integrand has a simple $e_0$-dependence of $e_0(1-e^2_0)^{-1/3}$ that can be integrated analytically.
Integrating~\eqref{eq:e0-integrand-GW} on the interval $[e_{0,{\rm th}},1)$ gives the GW spectrum in Eq.~\eqref{eq:OmegaGW_low}
\begin{align}
\Omega_{\rm GW,IR}\approx \dfrac{85^{\frac{2}{5}}\pi^{\frac{8}{15}}[\Gamma(\frac{7}{3})]^{\frac{3}{5}}f^{\frac{6}{5}}\left(G^{14}\GpoG^{17}H^4_0\mdm^{14}(1+z_{\rm eq})\Omega^{17}_{\rm DM}\right)^{\frac{1}{15}}}{2^{\frac{77}{15}}3^{\frac{4}{15}}c^{\frac{4}{5}}_1c^{\frac{4}{5}}_2 (\GpoG-1)^{\frac{2}{5}}}f^{2/3}_{\rm GW}\,,\nonumber
\end{align}
which we refer to as the IR piece of the spectrum.
The absence of $\rhodm$ in the expression (as opposed to the UV piece where it is present) is a result of the approximation $\Gamma(\frac{7}{3},B\uu\tau^{1/4}_{\rm min})\approx\Gamma(\frac{7}{3})$.
A sanity check of the condition $B\uu\tau^{1/4}_{\rm min} \ll 1$ using the $e_{0,{\rm th}}$ obtained in~\eqref{eq:e0th} roughly requires
\begin{align}
\left(\dfrac{\fGW}{1\uu{\rm Hz}}\right)\left(\dfrac{f}{0.01}\right)^{3/5}\left(\dfrac{\GpoG}{10^4}\right)^{1/5}\left(\dfrac{\mdm}{M_\odot}\right)^{-1/5}\left(\dfrac{\rhodm}{(1\uu\GeV)^4}\right)^{-1/2}\lesssim 2\times 10^{12}\,.
\end{align}

\begin{figure}
    \centering
    \includegraphics[width=0.6\textwidth]{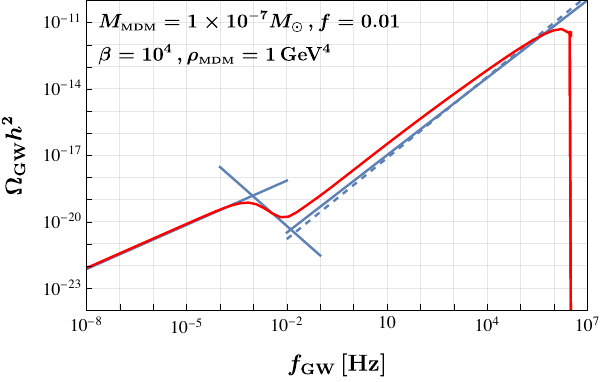}
    \caption{Comparison between our estimated power-law spectra (blue) and numeric results (red). The dashed line assumes radiation domination. 
    }
    \label{fig:OmegaGW_cmp}
\end{figure}

A comparison between our calculated power-law spectrum and the numeric results is shown in Fig.~\ref{fig:OmegaGW_cmp} (as well as Fig.~\ref{fig:spectra_shape}).
The analytic estimate is computed in the following way. The IR piece (\ref{eq:OmegaGW_low}) and the middle piece (\ref{eq:OmegaGW_middle}) of the spectrum are generated at early times. Both involve a competition between the GW and EM emission, which dominates for the IR and middle piece, respectively. This competition is in the denominator of $dE_{\rm GW}/d\fGWs$, therefore the smaller of the two spectra should be used for the early-time contribution. Meanwhile, the UV piece (\ref{eq:OmegaGW_high}) is generated at late times. Adding the early- and late-time contributions together,
\begin{align}
\Omega_{\rm GW}(\fGW)\approx {\rm Min}\big\{\Omega_{\rm GW,IR}(\fGW),\;\Omega_{\rm GW,middle}(\fGW)\big\}+\Omega_{\rm GW,UV}(\fGW)\,.
\end{align}

Some final comments on the competition between the GW process and the EM process in the denominator of $dE/df_{{\rm GW},s}$. It can be seen from Eq.~\eqref{eq:dEdfs_complete} that the GW process has a slightly larger power in $\mdm$, and hence dominates a larger parameter space for heavier MDM. This explains why the middle piece is absent for the spectra of larger $\mdm$.
The same logic suggests that the GW process is more favored when $\tau_{\rm min}$ is smaller, which could explain why the middle piece is less significant when $\rhodm$ is larger, as seen in the two benchmark points in Fig.~\ref{fig:spectra_shape}.

\section{GW emission at higher harmonics}\label{app:higher_harmonics}

As noted in the main text, our calculation effectively assumes that all the GW emission is through the  $\fGWs=2\uu f_{\rm orb}$ harmonic, while for a binary with an eccentric orbit the GWs are emitted at all harmonics of the orbital frequency, \ie, $\fGWs=n f_{\rm orb}$ where $n$ is a positive integer.
We have checked that the $n\neq 2$ harmonic channels are not the main contributors to $\Omega_{\rm GW}$.
Here we briefly review the formalism to include all harmonics, and then present the results.

The building block of the formalism is the Fourier analysis of the Kepler motion (see, \eg, Section 4.5 of~\cite{Maggiore:08textbook} for a review).
Since we treat the mediator as massless and hence the binary orbital motion as Kepler-like up to a redefinition of $G$ [see Eq.~\eqref{eq:kepler}], the mode decomposition should remain the same.
Therefore~\cite{Peters:1963ux,Enoki:2006kj},
\begin{align}
\dfrac{dE_{\rm GW}}{dt}=&\dfrac{32 G G'^3 \eta^2 m^5}{5 a^5} \dfrac{1 + \frac{73}{24}e^2 + \frac{37}{96}e^4}{(1-e^2)^{7/2}}=\dfrac{32 G G'^3 \eta^2 m^5}{5 a^5}\sum_n g(n,e)\,,
\end{align}
with 
\begin{align}
g(n,e)=&\dfrac{n^4}{32}\Bigg\{\left[J_{n-2}(ne)-2eJ_{n-1}(ne)+\dfrac{2}{n}J_n(ne)+2eJ_{n+1}(ne)-J_{n+2}(ne)\right]^2\nonumber\\
&+(1-e^2)\left[J_{n-2}(ne)-2eJ_n(ne)+J_{n+2}(ne)\right]^2+\dfrac{4}{3n^2}J^2_n(ne)\Bigg\}\,,
\end{align}
where $J_n(z)$ is the $n$-th order Bessel function.

\begin{figure}
    \centering
    \includegraphics[width=0.6\textwidth]{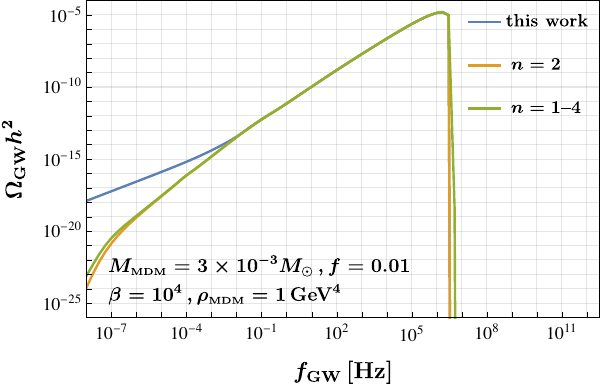}
    \caption{Comparison between the $\Omega_{\rm GW}h^2$ calculated using our approach and using the harmonic expansion approach in Eq.~\eqref{eq:dEGWdfGWs_n}. Our approach agrees very well with the $n=2$ harmonics which dominates the spectrum in the large $\fGW$ regime where EM emission dominates, and can further automatically incorporate the GW emission domination regime at small $\fGW$.}
    \label{fig:higher_harmonics}
\end{figure}

Then, 
\begin{align}
\dfrac{d^2E_{\rm GW}}{dt\uu d\fGWs}=&\dfrac{32 G G'^3 \eta^2 \mtot^5}{5 a^5}\sum_n g(n,e)\delta(\fGWs-nf_{\rm orb})\,.
\end{align}
After time integration,
\begin{align}
\dfrac{dE_{\rm GW}}{d\fGWs}=&\sum_n \int dt \dfrac{32 G G'^3 \eta^2 \mtot^5}{5 a^5}g(n,e)\delta(\fGWs-nf_{\rm orb})\nonumber\\
=&\sum_n \int \dfrac{de}{de/dt} \dfrac{32 G G'^3 \eta^2 \mtot^5}{5 a^5}g(n,e)\delta(\fGWs-nf_{\rm orb})\nonumber\\
=&\sum_n\int de \dfrac{32 G G'^3 \eta^2 \mtot^5}{5 a^5}\dfrac{a^3(1-e^2)^{3/2}}{4e G^2(\GpoG-1)\GpoG \mdm^2}\dfrac{g(n,e)}{n}\delta(f_{\rm orb}-\fGWs/n)\nonumber\\
=&\sum_n \dfrac{32 G G'^3 \eta^2 \mtot^5}{5 a^5}\dfrac{a^3(1-e^2)^{3/2}}{4e G^2(\GpoG-1)\GpoG \mdm^2}\dfrac{g(n,e)}{n}\dfrac{1}{\big\lvert\frac{df_{\rm orb}}{de}\big\vert_{e=e_n}}\label{eq:dEGWdfGWs_n}\\
\equiv&\sum_n\left(\dfrac{dE_{\rm GW}}{d\fGWs}\right)_n\,,
\end{align}
where $e=e_n$ is the root to the equation
\begin{align}\label{eq:e_n}
\dfrac{\fGWs}{n}=f_{\rm orb}=\dfrac{(G^\prime \mtot)^{1/2}}{2\pi a^{3/2}}=\dfrac{(G^\prime \mtot)^{1/2}}{2\pi}\left(\dfrac{e^{4/3}_0/(1-e^2_0)}{a_0 e^{4/3}/(1-e^2)}\right)^{3/2}\,.
\end{align}
The energy spectrum $(dE_{\rm GW}/d\fGWs)_n$, corresponding to the $dE_{\rm GW}/d\fGWs$ in Eq.~\eqref{eq:GW_spec2}, can be used to calculate the SGWB from the $n$-th order harmonic. Fig.~\ref{fig:higher_harmonics} shows a comparison between the calculated $\Omega_{\rm GW}h^2$ using the approach in the main text, along with the $n=2$ harmonics and the summed $n=$1--4 harmonics $\Omega_{\rm GW}h^2$ using Eq.~\eqref{eq:dEGWdfGWs_n}. One can see that indeed the $n=2$ harmonics dominates the contribution, and our approach agrees with the $n=2$ harmonics very well in the high $\fGW$ regime.
The deviation in the small $\fGW$ region is due to our explicitly using the orbital evolution results from pure EM emission in~\eqref{eq:dEGWdfGWs_n} and~\eqref{eq:e_n}, while in the small $\fGW$ region we know the GW emission process dominates the spectrum, as analyzed in Sec.~\ref{sec:SGWB_shape}.

\bibliography{DarkBinary}
\bibliographystyle{JHEP}

\end{document}